\documentclass[apj]{emulateapj}
\usepackage{apjfonts}

\shorttitle{Multifractal Analysis of Pulsar Timing Residuals}
\shortauthors{I.~ Eghdami, H.~ Panahi and S.~M.~S.~Movahed}
\usepackage{color}
\usepackage[colorlinks=true,
            linkcolor=black,
            urlcolor=black,
            citecolor=blue, breaklinks=true]{hyperref}
\usepackage{url}
\usepackage{breakurl}

\usepackage{amsmath}
\usepackage{amssymb}
\usepackage{gensymb}

\definecolor{mycolor}{rgb}{0.8,0.7,0.5}

\begin{document}

\title{Multifractal Analysis of Pulsar Timing Residuals: \\Assessment  of Gravitational Wave Detection}

\author{I.~ Eghdami\altaffilmark{1}, H.~ Panahi\altaffilmark{1} and S.~M.~S.~Movahed\altaffilmark{2,3}}
\affil{\scriptsize
$^{1}${Department of Physics, University of Guilan, Rasht 41635-1914, Iran}\\
$^{2}${Department of Physics, Shahid Beheshti University, Velenjak, Tehran 19839, Iran}\\
$^{3}$ School of Physics, Institute for Research in Fundamental Sciences (IPM), P. O. Box 19395-5531, Tehran, Iran}
\email{Corresponding author email: t-panahi@guilan.ac.ir}
\email{Corresponding author email: m.s.movahed@ipm.ir}
\begin{abstract}

We introduce a pipeline including multifractal detrended cross-correlation analysis (MF-DXA) modified by either singular value decomposition or the adaptive method to examine the statistical properties of the pulsar timing residual ($PTR$) induced by a gravitational wave (GW) signal. We propose a new algorithm, the so-called irregular-MF-DXA, to deal with irregular data sampling. Inspired by the quadrupolar nature of the spatial cross-correlation function of a gravitational wave background, a new cross-correlation function, $\bar{\sigma}_{\times}$, derived from irregular-MF-DXA will be introduced. We show that, this measure reveals the quadrupolar signature in the $PTRs$ induced by stochastic GWB.  We propose four strategies based on the $y$-intercept of fluctuation functions, the generalized Hurst exponent, and the width of the singularity spectrum to determine the dimensionless amplitude and power-law exponent of the characteristic strain spectrum as $\mathcal{H}_c(f)\sim\mathcal{A}_{yr}(f/f_{yr})^{\zeta}$ for stochastic GWB. Using the value of Hurst exponent, one can clarify the type of GWs. We apply our pipeline to explore 20 millisecond pulsars observed by Parkes Pulsar Timing Array. The computed scaling exponents confirm that all data are classified into a nonstationary class implying the universality feature. The value of the Hurst exponent is in the range  $H\in [0.56,0.87]$. The $q$-dependency of the generalized Hurst exponent demonstrates that the observed $PTRs$ have multifractal behavior, and the source of this multifractality is mainly attributed to the correlation of data which is another universality of the observed datasets. Multifractal analysis of available $PTRs$ datasets reveals an upper bound on the dimensionless amplitude of the GWB, $\mathcal{A}_{yr}< 2.0\times 10^{-15}$.

 \end{abstract}


\section{Introduction}
\label{sec:intro}

Pulsar timing has received extensive attention for astrophysical interests due to possessing a stable rotational mechanism \citep{Verbiest,lorimer}. The pulsar timing residual ($PTR$)  which is an important observable, is defined by the difference between the measured time of arrival (TOA) and those anticipated by a timing model \citep{Verbiest}.  The observed $PTR$ is a precise indicator to elucidate some interesting physical properties of pulsars and other cosmological and astrophysical foreground processes \citep{PPTA}. The influence of unknown physical phenomena on the variation of the pulsar's spin and spin-down and the presence of foreground effects impose randomness on the $PTR$. Therefore, the $PTR$ is categorized in a $(1+1)$-dimensional stochastic process (where only one of the parameters is independent, while the other parameter is represented as a function of the independent parameter). Therefore, the stochastic nature of the data requires implying robust methods to extract reliable information from the $PTR$.

Millisecond pulsars (MSPs) were first suggested as detectors of gravitational waves (GWs) by \cite{Sazhin78} and  \cite{Detweiler79}  because of the high stability and predictability of their rotational behavior (see also \citep{Hobbs_Tempo2}). Indeed, GWs can be produced by different mechanisms ranging from the early epoch to the present era. Continuous wave sources \citep{peters1964},  burst sources \citep{thorne1976,Damour01, kocsis06} and stochastic backgrounds \citep{enoki07, Maggiore} are the most well-known classes among the GW sources. As an illustration,  we refer to relic GWs, including GWs by cosmic strings and primordial perturbations \citep{Maggiore,Hobbs_GW,damour05,psh10}. The GWs are also produced in the formation of supermassive black holes and binary black hole mergers \citep{Rajagopal,Taylor:2012db}. Recently, the GWs of black hole mergers have been detected by LIGO instruments, which can be an evidence of dark matter in the early universe or can correspond to the binary black hole of stellar origin \citep{Mandic}.
Other classes of GWs include the continuous, inspiral, burst, and stochastic types of GWs \citep{Meadors,Coyne}.  Many approaches have been  proposed and utilized during past decades to detect mentioned types of GWs \citep{Coyne,Pai,zhu14,Jenet}. The very low amplitude of GWs, different sources and mechanisms for GW production on one hand, and  the extended range of frequency on the other hand lead to introducing various indirect approaches such as predictions of energy loss due to GW emission \citep{Taylor} and direct approaches such as detecting the effect of GWs on pulsar timing residuals \citep{Jenet}. The two main methods for detection of GWs are known as interferometers (such as LISA and LIGO) and pulsar timing arrays \citep{Roebber}.

%(such as EPTA and PPTA)

For the frequency interval $\nu\in [10^{-8},10^{-6}]$, there are several pulsar timing array projects that observe the imprint of GWs using pulsar timing detectors \citep{jenet06}. In the context of pulsar timing array approach, some famous projects have been proposed, namely the Parkes Pulsar Timing Array (PPTA) \citep{PPTA,man08,hob13}, the European Pulsar Timing Array (EPTA) \citep{jan08,kra13}, the North American Nanohertz Observatory for Gravitational Waves (NANOGrav) \citep{dem09,mc13} and International Pulsar Timing Array (IPTA) \citep{Verbiest}.
The Square Kilometer Array (SKA) \citep{cor04,laz13} radio telescope would further improve the sensitivity of pulsar timing measurements to detect GWs. For a recent and more complete discussions on various experiments and methods to detect GWs, see  \cite{zhu15,zhu16,and15,ellis14,Manchester10,George:2017pmj} .

A pulsar timing array utilizing the Parkes radio telescope in Australia is an experiment to detect GWs by observing 20 bright MSPs \citep{PPTA}.
The PPTA observations must be continued more than 5 yr in order to achieve a precision of 100~ns.
Since pulsar timing residuals are induced by GWs \citep{Jenet}, therefore some authors used the statistical correlation of pulsars timing residuals to evaluate their capability of detecting GWs \citep{Hellings,Jenet}.

In order to elaborate the benchmark of different types of GWs based on $PTR$ datasets,  we should consider four aspects. First of all, various sources of GWs encourage us to find deep insight regarding the properties of GWs emitted by different sources. Second, we should examine the performance of different statistical methods and their sensitivities. Third, we should take into account the quadrupolar signature of GWs. Finally, the upper limit on the amplitude of the gravitational wave background (GWB) should be computed.
The main method to detect a stochastic GW employing pulsar timing arrays is to search for a correlation between $PTR$s and compare it with quadrupole spatial cross-correlation calculated by {\it Hellings and Downs} \citep{Hellings,Jenet,Haasteren09,Haasteren11}.
If such a signature is not detected, one can set an upper bound on the amplitude of GWs using frequentist or Bayesian approaches. The upper limits provided by \cite{Haasteren11} and \cite{Shannon15} for a stochastic background
produced by supermassive black hole binaries is $\mathcal{A}_{yr}\leq
6\times10^{-15}$ and $\mathcal{A}_{yr}\leq 1\times10^{-15}$, respectively,
where the latter is the lowest claimed upper limit so
far. In addition, \cite{Demorest13} have used the high-precision pulsar timing data recorded as part of the NANOGrav project and finally provided  an upper  limit  on  the power-spectrum amplitude of the nHz-frequency stochastic supermassive black hole GWB.

Pulsar datasets are manipulated by trends and noises. Statistical models for noises, trends, and signals play crucial roles in any parametric GW detection approaches. Subsequently, it is necessary to implement robust and novel methods for removing destructive effects from desired parts of signals.

Our work in this paper has the following advantages and novelties.

 i) Inspired by the  properties of a self-similar process characterized by a scaling exponent called "Hurst exponent"
\citep{Hurst51,Hurst51-1,Hurst51-2,Hurst51-3}, for the first time, we have used Multifractal Detrended Fluctuation Analysis (MF-DFA) \citep{Peng95,Kantelhardt}, Multi-Fractal Detrended Moving Average Analysis (MF-DMA) \citep{alessio2002second,carbone2004analysis,arianos2007detrending,Gu.g,Shao.y}
 and Multifractal Detrended Cross-correlation Analysis (MF-DXA) \citep{DCCA,mf-dxa} methods to analyze the observed (including 20 MSPs inferred from \citep{PPTA}) and simulated pulsar timing residuals induced by GW signals (simulated by the TEMPO2 software package \cite{tempo2_I}).  We will evaluate the multiscaling behavior of the underlying data from statistical point of view.

ii) We modify MF-DFA, MF-DMA and MF-DXA by additional denoising algorithms, namely Adaptive Detrending (AD) \citep{hu09} or Singular Value Decomposition (SVD)   \citep{golub,trend3-1,trend3} methods to exclude or at least to reduce the contribution of unknown trends and noises as much as possible.  These methods are used as precomplementary denoising procedures.

iii) The standard version of multifractal analysis is a reliable algorithm when the input is a regular sampling series. Observed $PTR$s are unevenly sampled data sets; we therefore modify parts of the MF-DXA algorithm and call it the irregular MF-DXA method. In addition, noise modeling can be revealed by multifractal analysis.

iv) We check the multifractal nature of  pulsar timing residuals. We also determine the sources of multifractality based on our statistical approaches.

v) According to the quadrupolar signature on the spatial cross-correlation function of {\it PTR}s, the detectability of the stochastic GWB is evaluated according to the MF-DXA of {\it PTR}s. The cross-correlation exponent will be determined. We also give a new spatial cross-correlation function   %generalized form for spatial-temporal cross-correlation function
for pulsar timing residues.

vi) We introduce some criteria not only for discrimination of the stochastic GWB footprint and single sources GWs on pulsar timing residuals but also for determining the dimensionless amplitude of GWB. Then, an upper bound on the amplitude of stochastic GWs will be computed. Finally, this view provides a new insight to use pulsar timing residuals for further astrophysical studies.

The rest of this paper is organized as follows. In section \ref{MFDFA},  we will explain MF-DFA, MF-DMA, and MF-DXA, dealing with irregular sampled data, AD, and SVD in detail. A new  measure for the spatial cross-correlation of $PTR$ is presented in this section. Noise and trend modeling and  posterior analysis to obtain scaling exponents are also discussed in section \ref{MFDFA}.   Section \ref{datadis} is devoted to the theoretical notions of the GWB and data description for observed as well as synthetic datasets. We will implement the multifractal methods on simulated  timing residuals series in section \ref{applyingmethod}. We will also study a new spatial cross-correlation function derived by the MF-DXA method in the search of the footprint of the stochastic GWB in the sensitive range of pulsar timing residual series.  Four strategies to reveal the imprint of GWs on the residual time series in a noiseless observation will also  be explained in section \ref{applyingmethod}. We will examine the multifractality of the observed pulsar timing residuals in section \ref{applyingmethod2}. In that section, we will also look for an upper bound on the amplitude of stochastic GWs using some observed $PTR$s. Section \ref{sec:discussion} is devoted to summary and conclusion.

\section{Methodology: Multifractal analysis}\label{MFDFA}
Nonstationary sources such as  trends and artificial noises usually
influence the observed time series. To infer reliable results, these
spurious effects should be well characterized and distinguished from
the intrinsic fluctuations. Concerning trends, \cite{wu07} stated
that, in principle, there is no universal definition for trends, and
any proper algorithm for evoking trends from underlying  series
should remove the contribution of trends without  destroying
fluctuations. One of the well-studied methods  for this purpose is
MF-DFA \citep{Peng95,Kantelhardt}, used in various areas, such as
economical time series
\citep{economics,economics-1,economics-2,economics-3,movahedstock17},
river flow \citep{sadeghriver,movahed_SVD}, sunspot fluctuations
\citep{sadeghsun,hu09}, cosmic microwave background radiations
\citep{sadeghcmb}, music \citep{jafarimusic1,jafarimusic1-1}, plasma
fluctuations \citep{movahedplasma}, identification of a defective single layer in two-Dimensional material \citep{shidpour18}, traffic jamming
\citep{Xiao-Yan}, image processing, medical measurements
\citep{soares1,soares2}, and astronomy \citep{Zunino}.
Cross-correlation has also been introduced and applied in some cases
\citep{DCCA,DCCA2,DCCA3,DCCA4,zebende11,zebende13,kris15}. The
MF-DXA examining higher-order detrended covariance was introduced by \cite{mf-dxa}.
Although the approaches in multifractal detrended analysis, such as
the MF-DFA and MF-DXA methods, diminish polynomial trends, previous
researche demonstrated that sinusoidal and power-law trends are not
completely removed \citep{kunhu,trend2}. Mentioned trends make some
crossovers in fluctuation functions
\citep{kunhu,trend2,physa,trend3,SVD,trend3-1}. Several robust
methods have been proposed to eliminate cross-overs produced by
sinusoidal and power-law trends: Fourier Detrended Fluctuations
Analysis (F-DFA) \citep{f-dfa,SVD},  Singular Value Decomposition
(SVD) \citep{golub,trend3-1,trend3}, Adaptive Detrending method (AD)
\citep{hu09}, and Empirical Mode Decomposition (EMD)\citep{hu98}. In
this work, we implement the AD and SVD methods to reduce the contribution of noise and magnify the effect of GWs in our results for further cleaning preprocessors.

\subsection {Multifractal-based analysis}

Finding scaling exponents in the context of auto-correlation and cross-correlation analysis has many inaccuracies due to nonstationarity,  noises, and undesired trends. To resolve the mentioned difficulties, a well-known method based on decomposing the original signal into its positive and negative fluctuation components has been proposed by \cite{jun06}. Motivated by such a  decomposition method, Podobnik et al. introduced the cross-correlation between two non-stationary fluctuations by means of the DFA method \citep{DCCA}. A modification of detrended cross-correlation analysis (DCCA) is known as MF-DXA was invented by \cite{mf-dxa}.
The pipeline of MF-DXA is considered as follows \citep{DCCA,mf-dxa}\footnote{If both signals are identical, we have the MF-DFA/MF-DMA method.}.

$(1)$: We consider two typical $PTR$ series named by  $PTR_{a}$ and $PTR_{b}$, located at $\hat{n}_a$ and $\hat{n}_b$ with respect to the line of sight, respectively, as the input data sets to study their mutual multifractal property:
\begin{eqnarray}\label{00}
&&{\it PTR_{a}}(i), \quad {\it PTR_{b}}(i), \qquad i=1,...,N
\end{eqnarray}
The pulsar timing observations are almost unevenly sampled datasets.  We need equidistant sampling series. A trivial but not essentially optimum way is to interpolate between two successive data.  Different methods to reconstruct regular series  will be explained in subsection \ref{datairregular}. Therefore, here we assume that the input data are regular and ready for further tasks. Moreover, the observed data have variable error bars, and, to take into account heteroskedasticity, we use error-propagation formalism in all statistical analysis, such as averaging, fitting, and computing fluctuation functions  throughout this paper.

$(2)$: To magnify the hidden self-similarity property, we make profile series according to:
\begin{eqnarray}\label{profile22}
X_{\diamond}(j)&=&\sum_{i=1}^{j} [{\it PTR_{\diamond}}(i)-\langle {\it PTR_{\diamond}}\rangle], \quad j=1,...,N
\end{eqnarray}
Here the subscript $\diamond$ can be replaced by "$a$" or "$b$".

$(3-a)$: The above profile series must be divided into $N_{s}=\mathrm{int}(N/s)$ nonoverlapping segments of length $s$. The range of nonoverlapping window values is $N_{s}\in[N_s^{\rm min},N_s^{\rm max}]$. To take into account the remaining unused part of the data from the opposite end of the data, the enumeration must to be repeated from the mentioned part. In this case, we will have $2N_s$ segments. In the framework of the MF-DCCA method, we should compute the following fluctuation function in each segment as follows:
\begin{eqnarray}\label{flucdfa1}
{\mathcal{E}}_{\times}(s,\nu)=\frac{1}{s}\sum^{s}_{i=1}&&\left[ X_{a}(i+(\nu-1)s)-\tilde{X}_{a}^{(\nu)}(i)\right]\nonumber\\
&&\times \left[ X_{b}(i+(\nu-1)s)-\tilde{X}_{b}^{(\nu)}(i)\right]
\end{eqnarray}
for segments $ \nu=1,...,N_{s} $. For the opposite end, we have:
\begin{eqnarray}\label{flucdfa2}
{\mathcal{E}}_{\times}(s,\nu)=\frac{1}{s}\sum^{s}_{i=1}&&\left[ X_{a}\big(i+N-(\nu-N_{s})s\big)-{\tilde{X}}_{a}^{(\nu)}(i)\right]\nonumber\\
&&\times \left[ X_{b}\big(i+N-(\nu-N_{s})s\big)-{\tilde{X}}_{b}^{(\nu)}(i)\right]\nonumber\\
\end{eqnarray}
where $ \nu=N_s+1,\cdot\cdot\cdot,2 N_{s} $ and $\tilde{X}_{\diamond}^{(\nu)}(i)$ is a weighted fitting polynomial function in the $\nu$th segment with an arbitrary order describing the local trend for data with variable error bars. Usually a linear function for modeling local trends is considered \citep{bunde00}.
The MF-DCCA$m$ denotes that the order of the polynomial function used in the MF-DCCA is "$m$''. Throughout this paper, we take $m=1$ unless stated otherwise. To reduce the statistical uncertainties in the computed fluctuation functions, we set $s>m+2$ \citep{Kantelhardt}. On the other hand, this method becomes unreliable for very large window sizes, i.e. $s>\frac{N}{4}$. There is a discontinuity for fitting a polynomial at the boundary of each partition in the MF-DCCA method; to resolve this discrepancy, MF-DMA has been introduced \citep{alessio2002second,carbone2004analysis,arianos2007detrending,Gu.g,Shao.y}. Accordingly, instead of doing item $(3-a)$, we carry out the following procedure:

$(3-b)$:  For each moving window with size $s$, we calculate the moving average function:
\begin{equation}
\widetilde{X_{\diamond}(j)}=\frac{1}{s}\sum_{k=-s_1}^{s_2}X_{\diamond}(j-k)
\end{equation}
where $s_1=\lfloor(s-1)\theta\rfloor$ and $s_2=\lceil(s-1)(1-\theta)\rceil$. The symbol $\lfloor a \rfloor$ represents the largest integer value not greater than $a$ and $\lceil a\rceil$ is devoted to the smallest integer value not smaller than $a$. In the above equation, $\theta$ plays a crucial role. The $\theta=0$ refers  to the backward moving average, while $\theta=1$ is the so-called forward moving average; finally $\theta=0.5$ is related to
the centered moving average \citep{xu05,Gu.g}. Therefore, detrended data are constructed by subtracting the calculated moving average function from the cumulative series, $X_{\diamond}$ as:
\begin{eqnarray}
\varepsilon_{X_{\diamond}}(i)= X_{\diamond}(i)-\widetilde{X_{\diamond}(i)}
\end{eqnarray}
where $s-s_1 \leq i\leq N-s_1$.  Now $\varepsilon_{X_{\diamond}}(i)$ values are divided into $N_{s}={\rm int}[N/s]$ nonoverlapping windows with the same size of $s$ and we calculate the fluctuation function:
\begin{equation}\label{flucdma}
\mathcal{E}_{\times}(s,\nu) = \frac{1}{s}\sum_{i=1}^{s}\varepsilon_{X_{a}}(i+({\nu}-1)s)\times \varepsilon_{X_{b}}(i+({\nu}-1)s)
\end{equation}

$(4)$: Using Eqs. (\ref{flucdfa1}) and (\ref{flucdfa2}) for the MF-DCCA (MF-DFA) based method \citep{peng92,peng94,bul95,Kantelhardt,sho12} and Eq. (\ref{flucdma}) for the MF-DMA algorithm, the corresponding $q$th-order fluctuation function can be computed by:
\begin{equation}
\label{fluc1DXA}
{\mathcal{F}}_{\times}(q,s)=\left( \frac{1}{ 2N_{s}}\sum^{ 2N_{s}}_{\nu=1}\left |{\mathcal{E}}_{\times}(s,\nu)\right|^{q/2}\right)^{1/q}
\end{equation}
For $q=0$, we have:
\begin{equation}
\label{fluc2DXA}
{\mathcal{F}}_{\times}(0,s)=\exp\left(   \frac{1}{4 N_{s}}\sum^{ 2N_{s}}_{\nu=1}\ln |{\mathcal{E}_{\times}}(s,\nu)| \right)
\end{equation}

$(5)$: The scaling behavior of the fluctuation function according to:
\begin{equation}
\label{eq: exponentdxa}
{\mathcal{F}}_{\times}(q,s)\sim s^{h_{\times}(q)}
\end{equation}
gives the cross-correlation exponent $h_{\times}(q)$. The $q$-parameter enables us to quantify the contribution of different values of fluctuation functions in Eqs. (\ref{fluc1DXA}) and (\ref{fluc2DXA}).  The small fluctuations play a major role in summation for $q<1$, while large fluctuations become dominant for $q\ge1$.  We emphasize that for heteroskedastic data, the summation in Eqs. (\ref{fluc1DXA}) and (\ref{fluc2DXA}) should incorporate variable errorbars, and weighted fitting polynomials must be considered. It turns out that for $a=b$, the usual generalized Hurst exponent, $h(q)$, is retrieved. In this case we have:
\begin{equation}
\label{eq: fluctuation function}
{\mathcal{F}}_{q}(s)= \mathcal{G}_{h(q)}s^{h(q)}
\end{equation}
for $q=2$, the $\mathcal{G}$ is
\begin{eqnarray}\label{eq:g}
\mathcal{G}&=&\frac{\sigma^2}{2H+1}-\frac{4\sigma^2}{2H+2}+3\sigma^2\left ( \frac{2}{H+1}-\frac{1}{2H+1}\right )\nonumber\\
&&-\frac{3\sigma^2}{H+1}\left (1-\frac{1}{(H+1)(2H+1)}\right)
\end{eqnarray}
and $\sigma^2=\langle PTR^2\rangle $ for zero mean data. Any $q$-dependency of $h(q)$, confirms that the underlying data set is a multifractal process. For the class of the nonstationary series (corresponding to a fractional Brownian motion; fBm) the exponent derived by using MF-DFA is $h(q=2)>1$. Therefore, in this case, the Hurst exponent is given by $H=h(q=2)-1$. In the stationary case, $h(q=2)<1$ (corresponding to a fractional Gaussian noise; fGn) and $H=h(q=2)$. For completely stationary random data, $H=0.5$, while for a persistent data set,  $0.5<H<1.0$. For an anticorrelated data set, $H<0.5$ \citep{taqq95,peng94,ossad95}. When the Hurst exponent is determined, the scaling exponents of autocorrelation for an fGn process read as $\mathcal{C}(\tau)=\langle x(t)x(t+\tau)\rangle \sim \tau^{-\gamma}$ for $\tau\gg0$ with $\gamma=2-2H$, while for a fBm signal, we have $\mathcal{C}(t_i,t_j)=\langle x(t_i)x(t_j)\rangle \sim t_i^{-\gamma}+t_j^{-\gamma}-|t_i-t_j|^{-\gamma}$ for $|t_i-t_j|\gg0$ with $\gamma=-2H$. The associated power spectrum is $S(f)\sim f^{-\beta}$ with $\beta = 2H-1$ and $\beta =2H+1$ for the fGn and fBm processes, respectively. The relation between the generalized Hurst exponent and the scaling exponent of the partition function known as the multifractal scaling exponent based on the standard multifractal formalism becomes  \citep{Kantelhardt}:
\begin{equation}
\label{eq: tau and h}
\xi(q)=qh(q)-1
\end{equation}
For a monofractal data set, $\xi(q)$ is a linear function \citep{Kantelhardt}. The generalized multifractal dimension is also given by:
\begin{equation}\label{dq}
D(q)=\frac{\xi(q)}{q-1}=\frac{qh(q)-1}{q-1}
\end{equation}
where $D(q = 0) = D_f$ is the fractal dimension of the time series and $D(q = 1)$ is related to the so-called
entropy of the underlying system \citep{Thomas86}. A more complete quantitative measure of multifractality is the singularity spectrum and indicates how the box probability of  standard multifractal formalism behaves at small scales. It is defined by the Legendre transformation of $\xi(q)$ as \citep{feder}:
\begin{equation}
f(\alpha)=\alpha q-\xi(q)
\label{eq:f(alpha)}
\end{equation}
and the H${\rm \ddot{o}}$lder exponent is $\alpha\equiv d\xi(q)/dq$. In the case of multifractality, a spectrum of  the H${\rm \ddot{o}}$lder exponent  is obtained instead of a single exponent. The domain of the H${\rm \ddot{o}}$lder spectrum, $\alpha \in
[\alpha_{\rm min},\alpha_{\rm max}]$, becomes \citep{muzy94,arneodo94}:
\begin{equation}
\alpha_{\rm min}= \lim_{q\to +\infty}\frac{\partial \xi(q)}{\partial q},\quad   \alpha_{\rm max}= \lim_{q\to -\infty}\frac{\partial \xi(q)}{\partial q}
\label{alpha}
\end{equation}
Subsequently, the width $\Delta \alpha\equiv \alpha_{\rm max}-\alpha_{\rm min}$ is a reliable measure for quantifying the multifractal nature of the underlying data. The higher value of $\Delta \alpha$ is associated with the higher multifractal nature reflecting the complexity of the signal. As other complexity measures, one can point to the $q$-order  Lyapunov exponent \citep{eckmann86}, and the Lempel-Ziv complexity \citep{lempleziv76}.
Inspired by the common cross-correlation definition, relying on Eq. (\ref{flucdma}), we define the new cross-correlation function \citep{zebende11,zebende13}:
  \begin{equation}\label{eq:newsigma0}
 \sigma_{\times}(\Theta_{ab})\equiv {\sum}_{s}\left(\frac{\sum^{ 2N_{s}}_{\nu=1}{\mathcal{E}}_{\times}(s,\nu)}{\sqrt{\left[\sum^{ 2N_{s}}_{\nu=1}{\mathcal{E}}_{a}(s,\nu)\right]\left[\sum^{ 2N_{s}}_{\nu=1}{\mathcal{E}}_{b}(s,\nu)\right]}}\right)
\end{equation}
here $\Theta_{ab}=\arccos|\hat{n}_a.\hat{n}_b|$.
Averaging on all available pairs separated by $\Theta$ leads to:
\begin{equation}\label{eq:newsigma}
\bar{\sigma}_{\times}(\Theta )=\frac{1}{4\pi}\int d\Omega {\sigma}_{\times}(\Theta_{ab})
\end{equation}
The $\bar{\sigma}_{\times}$ introduced by Eq. (\ref{eq:newsigma}) based on fluctuation functions computed in the context of detrended cross-correlation contains  the quadrupolar signature  if $PTR$s are modified by the GWB signal. Therefore, this is a new criterion that enables us to assess the footprint of GWs more precisely.

Now we turn to the spatial cross-correlation function for $PTR$s taking into account stationarity as:
 \begin{eqnarray}\label{eq:corrhel1}
 \mathcal{C}_{\times}(\Theta_{ab})&=&\langle PTR_a(t,\hat{n}_a)PTR_b(t,\hat{n}_b)\rangle_{t}
\end{eqnarray}
In the presence of an isotropic GWB, by averaging the cross-correlation on all available pairs separated by $\Theta$ leads to:
\begin{eqnarray}\label{corrhel}
\overline{\mathcal{C}}_{\times}(\Theta)&=&\langle \ \mathcal{C}_{\times}(\Theta_{ab})\rangle_{\rm pairs} \sim\overline{\Gamma}(\Theta)
\end{eqnarray}
The $\overline{\Gamma}(\Theta)$ is given by the {\it Hellings and Downs} equation \citep{Hellings,Jenet}:
\begin{eqnarray}\label{eq:gamma}
\overline{\Gamma}(\Theta)=\frac{3}{2}\psi\ln(\psi)-\frac{\psi}{4}+\frac{1}{2}
\end{eqnarray}
where $\psi \equiv {[1-\cos(\Theta)]}/{2}$. We should notice that the {\it Hellings and Downs} curve is only a function of the angular separation between pulsar pairs separated by $\Theta$, and it is independent of the frequency \citep{Romano}.

The new cross-correlation coefficient defined by Eq. (\ref{eq:newsigma}) is related to the traditional cross-correlation $\mathcal{C}_{\times}$ in a complex way, and the relation is not analytically tractable without any approximation, and we will evaluate it numerically in the next section. However, according to Eq. (\ref{flucdma}), the mapping between $\mathcal{C}_{\times}$ and $\sigma_{\times}$ does not change the sign of $\sigma_{\times}$. Thus, the quadrupolar signature of the {\it Hellings and Downs} function is preserved. It is worth mentioning that, besides probable GW signal superimposed in the $PTR$s, the following fluctuations can be existed in the recorded data: the correlated red (fractal) noise; clock errors, which are the same in all pulsars (i.e., monopolar); and ephemeris errors (which are dipolar).  There are no known noise sources other than GWs that are quadrupolar \citep{Tiburzi}.

 Applying MF-DXA on $PTR$s determines the value of the temporal scaling exponent, $h_{\times}$.  We expect to find constant $h_{\times}(q)$ with respect to different separation angles ($\Theta$) for an isotropic GWB, while for the other local source of GWs, the $h_{\times}(q)$ depends on $\Theta_{ab}$ in an arbitrary manner.

\subsection{Dealing with irregularly sampled data}\label{datairregular}
The pulsar timing observations are unevenly sampled; i.e. they are not
a set of equidistant sampling values, and the underlying series is
nonuniform, requiring some sort of interpolation technique. The
Lomb-Scargle periodogram proposed a least-squares pipeline to
resolve this problem  \citep{lomb76,scargle82}. Radon
transformations have also been used for irregular sampling analysis
\citep{ronen91,dui991,dui992}; see also \citep{gulati09} and
references therein. Extrapolation of irregularly recorded data onto
a regular grid was introduced by \cite{freg06}. For
constructing Fourier expansion,  nonuniform discrete Fourier
transform was introduced by \cite{gulati09, Anholm}. A trivial
but not necessarily optimum method with less computational burden is
to interpolate between two successive data points  in recorded
series. A more robust method is to apply kernel functions on the
irregular data, as (see also \citep{monaghon85}):
\begin{equation}\label{eq:kernel}
PTR_{reg}(t)=\int dt' PTR_{irre}(t')\mathcal{W}(t-t')
\end{equation}
where $PTR_{reg}$ and $PTR_{irre}$ are regular and nonuniform sampled data, respectively. Here $\mathcal{W}$ is a normalized window function. A typical functional form for this window function can be Gaussian.  In general, the choice of the window function, $\mathcal{W}$, depends on the smoothness, accuracy requirements, and computation efficiency \citep{monaghon85}.

Here we propose a new approach to find robust scaling properties for irregular sampled data. If there is no {\it a priori} information for the smoothing procedure, we suggest applying a gaussian kernel to the data followed by a linear interpolation to regularize datasets. Subsequently, we can construct the profile using such regular data (Eq.  (\ref{profile22})).  To reduce the contribution of artificial data points produced in this interpolation, we introduce tge irregular MF-DXA method. In this new algorithm, we modify the fluctuation function procedure given by Eqs. (\ref{flucdfa1}) and (\ref{flucdfa2})  for identical $PTR$s as:
\begin{equation}\label{fluc2new}
{\mathcal{E}}^{2}(s,\nu)=\frac{1}{s_{\nu}'(s)}\sum^{s_{\nu}'(s)}_{i=1}\left[ X(i+(\nu-1)s')-\tilde{X}_{\nu}(i)\right]^{2}
\end{equation}
In the above equation, only the data points recorded during observation in each segment with size $s$ will be considered for further computations. Therefore, the number of data in the $\nu$th window with size $s$ is represented  by $s_{\nu}'(s)$ which in general is not equal to $s$. Now Eq. (\ref{fluc1DXA}) becomes a weighted average:
\begin{equation}
\label{fluc1new}
{\mathcal{F}}_{q}(s)=\left(\frac{\sum^{2 N_{s}}_{\nu=1}\frac{\left[{\mathcal{E}}^{2}(s,\nu)\right]^{q/2}}{\sigma_{\mathcal{E}}^2(s,\nu,q)}}{ \sum^{2 N_{s}}_{\nu=1}\frac{1}{\sigma_{\mathcal{E}}^2(s,\nu,q)} } \right)^{1/q}
\end{equation}
where $\sigma_{\mathcal{E}}^2(s,\nu,q)$ is the variance of $\left[{\mathcal{E}}^{2}(s,\nu)\right]^{q/2}$.
We similarly replace the averaging procedure in any relevant parts with the weighted averaging.

Recently, \cite{ma10} showed that the global scaling exponents of
long-correlated signals remain unchanged for up to 90\% of data
loss, while for anticorrelated series, even less than 10\%  of data
loss creates a significant modification in the original scaling exponents.
This research shows that one can compute the scaling exponents for
long-range correlated irregularly sampled data points if one
regularizes the data set through linear interpolation and then
applies DFA. But for an anticorrelated signal, the DFA method does not
lead to reasonable results. Our new proposal demonstrates  that  for
synthetic series with known Hurst exponents, our modification leads
to more reliable estimations for scaling exponents, not only for
correlated series but also for anticorrelated datasets. Our
simulations show that the $PTR$ can be considered as long-range
correlated fluctuation. Therefore, our results are almost are not
affected by the type of regularization.
\subsection{SVD}
It is important to find trends and noise sectors in data analysis, especially in the astronomical data.
When we use MF-DFA, MF-DMA, and MF-DXA, an essential demand corresponding to presenting a scaling behavior must be satisfied, as represented by Eqs. (\ref{eq: exponentdxa}) and (\ref{eq: fluctuation function}).  In some cases, there exist one or more crossovers corresponding to different correlation behaviors of the pattern in various scales \citep{kunhu,trend2,physa,trend3,SVD,trend3-1}. The MF-DFA and MF-DXA methods cannot remove the effect of all undesired parts of the underlying signal; therefore, we implement complementary tasks to properly recover the scaling behavior of fluctuation functions properly and to obtain the reliable scaling exponents.
There are some preprocessing methods for denoising in the literature; for instance, The EMD method \citep{hu98}, the Fourier-detrended (Fourier-based filtering) method \citep{f-dfa,SVD}, the SVD method \citep{golub,trend3-1,trend3} and the AD algorithm \citep{hu09}. In this paper, we utilize the SVD method and AD algorithm.
The main part of the SVD method can be described in the following steps \citep{trend3,trend3-1,movahed_SVD}:

(I): Construct a matrix whose elements are $PTR$s in the following order: \begin{equation}
\mathbf{\Gamma}\equiv\left(
  \begin{array}{cccc}
    {\it PTR}_1 & {\it PTR}_{1+\tau} & ... & {\it PTR}_{1+N-(d-1)\tau-1} \\
   \vdots&\vdots&\vdots &\vdots \\
      {\it PTR}_i & {\it PTR}_{i+\tau} & ... & {\it PTR}_{i+N-(d-1)\tau-1} \\
   \vdots&\vdots&\vdots &\vdots \\
  {\it PTR}_d & {\it PTR}_{d+\tau} & ... & {\it PTR}_{d+N-(d-1)\tau-1} \\
  \end{array}
\right)\label{matrix1}
 \end{equation}
where  $d$ is the embedding dimension, $\tau$ is the time delay, and $1 \leq i\leq d$. Considering a time series of size $N$, the maximum value of the embedding dimension $d$ is equal to $d\leq N-(d-1)\tau+1$ \citep{trend3-1,SVD,shang09}.

(II): Decompose the matrix $\mathbf{\Gamma}$ to left (${\mathbf{U}}_{d\times d}$) and right (${\mathbf{V}}_{(N-(d-1)\tau)\times(N-(d-1)\tau)}$) orthogonal matrices:
\begin{equation}
\mathbf{\Gamma}={\mathbf{USV^{\dagger}}}
\end{equation}
where ${\mathbf{S}}_{d\times (N-(d-1)\tau)}$ is a diagonal matrix and its elements are the desired singular values. If we are interested in examining the fluctuations with high frequency, we should remove dominant wavelengths. In this case, for removing trends containing $p$ dominant wavelengths, we set $2p+1$ largest eigenvalues of matrix $\mathbf{S}$ to zero; therefore, long periods or short frequencies are eliminated. In other words, $p$ dominant eigenvalues and associated eigenvectors correspond to long wavelength (short-frequency part) subspace, while $d-p$ eigenvalues and the corresponding eigen-decomposed vectors represent short-wavelength (high-frequency part) subspace.

In this paper, we look for the footprint of GWs superimposed on the $PTR$s signals. As shown in Fig. \ref{fig:perfect_resi}, the GW part behaves as a dominant trend in $PTR$s; consequently, we essentially need to do denoising using the SVD method to magnify the contribution of superimposed GWs. To this end, we should remove small eigenvalues corresponding to a low-pass filter. In this paper, we eliminate the high-frequency part of the signal by keeping the $2p+1$ largest eigenvalues of the matrix $\mathbf{{S}}$.

Finally, the new eigenvalues matrix, $\mathbf{\tilde{S}}$, is determined. According to the filtered matrix, $\mathbf{\tilde{\Gamma}}={\mathbf{U\tilde{S}V^{\dagger}}}$, the cleaned time series is constructed by:
\begin{equation}\label{residusvd}
\widetilde{{\it PTR}}_{i+j-1}=\tilde{\Gamma}_{ij}.
\end{equation}
Here $1\le i\le d$ and $1\le j\le N-(d-1)\tau$. Now the cleaned $\widetilde{{\it PTR}}$ datasets will be used as input for the MF-DFA or MF-DXA discussed in previous subsections.
\subsection {AD algorithm}\label{method}

Another robust algorithm to examine trends is the AD method introduced by \cite{hu09}. The implementation of the AD algorithm is a complementary method for determining local and global trends. Therefore, after applying the AD method on observed pulsar timing series, the corresponding dominant trend output data will be used as an input for the MF-DFA or MF-DXA methods. The AD method includes the following steps \citep{hu09}. A discrete series, ${\it PTR}(i)$ with $i=1,\cdots,N$ is partitioned with overlapping windows of length
$2n+1$ and, accordingly,  each neighboring segment has $n+1$
overlapping points. An arbitrary polynomial ${\cal Y}$ is constructed in each window of
length $2n+1$.  In order to have the continuous trend function avoid a typical sharp jump in it, the following weighted function for the
overlapping part of the $\nu$th segment is considered \citep{hu09}:
\begin{eqnarray}\label{residuad}
{\cal Y}_\nu^{{\rm overlap}}(j)=\left(1-\frac{j-1}{n}\right){\cal Y}_\nu(j+n)+\frac{j-1}{n}
{\cal Y}_{\nu+1}(j)\;\nonumber\\
\end{eqnarray}
where $j=1,2,\cdots,n+1$. The two free parameters, namely $n$ and the order of the fitting
polynomial, should be determined properly \citep{hu09}.

The size of each segment was calculated by $2n+1= 2\times{\rm int}\left[(N-1)/(w_{\rm adaptive}+1)\right]+1$. It turns out that by increasing the value of $w_{\rm adaptive}$ and the order of the fitting polynomial, the fluctuations disappear, and, consequently,  the fluctuations are suppressed. For the nonoverlapping segments, the AD data are given by
${\it PTR}(i)-{\cal Y}_\nu (i)$, while for the overlap part it is 
${\it PTR}(i)-{\cal Y}_\nu^{\rm overlap}(i)$. Since the GW, as the dominant part of the signal, is our desired part of the signal, we instead use
$\widetilde{{\it PTR}}(i)={\cal Y}_\nu (i)$, while for the overlap part, we consider
$\widetilde{{\it PTR}}(i)={\cal Y}_\nu^{\rm overlap}(i)$. Now $\widetilde{{\it PTR}}(i)$ is used for further analysis in MF-DFA or MF-DXA.

\subsection{Trend and noise modeling}

In real observational data to carry out parametric detection,
reliable statistical models of the noise and signal should be well
established. A proposal for noise modeling is based on the denoiseing
procedure carried out by the SVD or AD algorithms. Previously, we were
interested in removing the contribution of undesired noise modulated
on real data. Now we  concentrate on the $\widetilde{{\it PTR}}$
given by Eq. (\ref{residusvd}) in the context of SVD analysis as a
model of trends and  $PTR-\widetilde{{\it PTR}}$ for noise. Also, if we use the AD approach, the global variation part of the signal corresponds to both $\mathcal{Y}$ and
$\mathcal{Y}^{\rm overlapp}$ (Eq. (\ref{residuad})). For the noise part, we should consider ${\it PTR}(i)-{\cal Y}_\nu (i)$, while for the overlap part, it is ${\it PTR}(i)-{\cal Y}_\nu^{\rm overlap}(i)$. Therefore, SVD or AD, as well as the internal part of the MF-DFA and MF-DXA algorithms, are
able to give a robust model for trends and noise. Also, extracting
intrinsic functions based on EMD can be a good proposal for this
purpose \citep{hu98}.

\subsection{Posterior Analysis}

In this paper, we turn to Bayesian statistics \citep{fab04} to compute the reliable value of the generalized Hurst exponent (Eqs. (\ref{eq: exponentdxa}) and (\ref{eq: fluctuation function})). Let $\{{\mathcal {D}}\}:\{{\mathcal F}_q(s)\}$ and $\{\Upsilon\} : \{h(q)\}$ represent the measurements and model parameters, respectively. The posterior function is defined by:
\begin{equation}\label{posterior}
{\mathcal P}(\Upsilon |{\mathcal {D}})=\frac{{\mathcal{L}}({\mathcal {D}}|\Upsilon){\mathcal P}(\Upsilon)}{\int
{\mathcal{L}}({\mathcal {D}}|\Upsilon){\mathcal P}(\Upsilon)d\Upsilon}
\end{equation}
where $\mathcal{L}$ is the likelihood and $\mathcal{P}(\Upsilon)$ is the prior probability function including all information concerning model parameters. Here we adopt the top-hat function for ${\mathcal{P}}(h(q))$ in the interval $h(q)\in[0,4]$. According to the central limit theorem, the functional form of likelihood becomes multivariate Gaussian, i.e. ${\mathcal{L}}({\mathcal {D}}|\Upsilon)\sim \exp(-\chi^2/2)$.
The $\chi^2$ for determining the best-fit value for the scaling exponent coordinated by multifractal formalism reads as:
\begin{equation}
\chi^2(\Upsilon)\equiv \Delta^{\dag}.{C}^{-1}.\Delta
\end{equation}
where $\Delta\equiv [{\mathcal F}_q^{{\rm obs.}}-{\mathcal F}_q^{{\rm the.}}]$ and ${C}$ is the covariance matrix. The ${\mathcal F}_q^{{\rm obs.}}(s)$ and ${\mathcal F}_q^{{\rm the.}}(s;h(q))$ are
fluctuation functions computed directly from the data and determined by Eqs.  (\ref{eq: exponentdxa}) or (\ref{eq: fluctuation function}), respectively.
In the case of the diagonal covariance matrix, the $\chi^2$ becomes:
\begin{equation}
\chi^2(h(q))=\sum_{s=s_{\rm min}}^{s=s_{\rm max}}\frac{[{\mathcal F}_q^{{\rm obs.}}(s)-{\mathcal F}_q^{{\rm the.}}(s;h(q))]^2}{\sigma_{{\rm obs.}}^2(s)}
\end{equation}
Here $\sigma_{{\rm obs.}}(s)=\left\langle \left[\delta\mathcal{F}_q^{{\rm obs.}}(s)\right]^2\right\rangle$, which is related to the diagonal elements of ${C}$ and can be computed using a standard statistical error propagator from primary uncertainties on $PTR$ datasets (Eq. (\ref{00}) to Eqs. (\ref{fluc1DXA}) and (\ref{fluc2DXA})). The $1\sigma$ error bar of $h(q)$ is determined by:
\begin{equation}
68.3\%=\int_{-\sigma^{-}_{h(q)}}^{+\sigma^{+}_{h(q)}}{\mathcal{L}}({\mathcal F}_q(s)|h(q))dh(q)
\end{equation}
Subsequently, we report the best value of the scaling exponent at a $1\sigma$
confidence interval as $h(q)_{-\sigma^-_{h(q)}}^{+\sigma^+_{h(q)}}$.
\begin{figure}[t]
\centering
\includegraphics[scale=0.35]{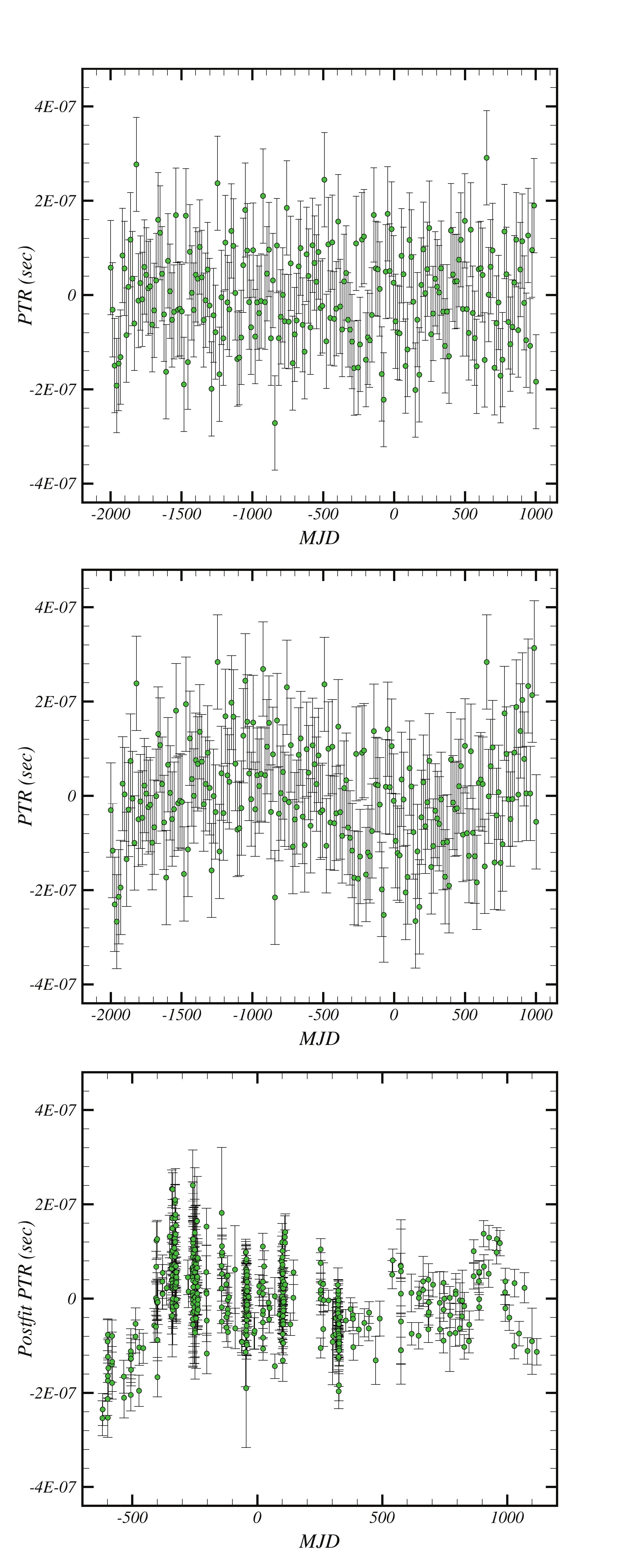}
\caption{The upper panel corresponds to a pure simulated timing residual. The middle panel shows a synthetic pure timing residual induced by the GWB with a dimensionless amplitude of $\mathcal{A}_{yr}=10^{-15}$. Here we take $\zeta=-2/3$. The lower part shows the observed pulsar timing residual of PSR J0437-4715 from the PPTA project.}
 \label{fig:perfect_resi}
 \end{figure}
\begin{figure}
\centering
\includegraphics[scale=0.32]{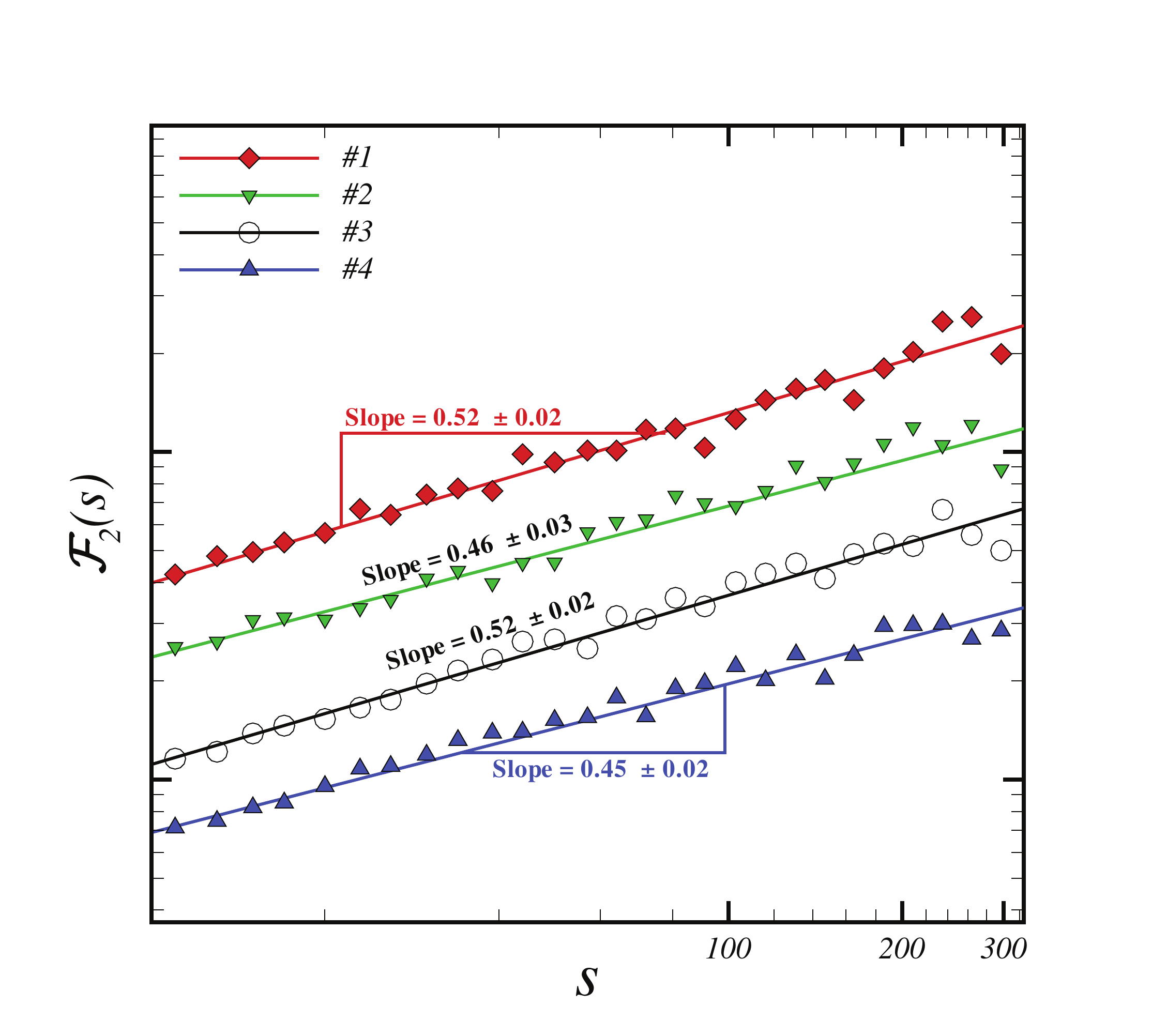}
\caption{ Log-log plot of $\mathcal{F}_2(s)$ versus $s$ computed by DMA with $\theta=0.0$ for various simulated pure {\it PTR}s. To make more sense, we shifted $\mathcal{F}_2$ vertically for different series. As we expect, the value of the Hurst exponent for all datasets is consistent with completely random series.}
\label{fspure}
\end{figure}

\section{Data Description}\label{datadis}
In this section, we will describe theoretical models for GW signals.
The observational datasets, synthetic series for pure timing residuals, and GWs,  in order to examine the multiscaling behavior of  $PTR$s as an indicator of GWs, will be described in this section.

\subsection{Theoretical notions of the GWB on {\it PTR}s}
The potential sources of GWs could be massive accelerated objects \citep{Rajagopal,Taylor:2012db}, burst sources \citep{thorne1976,Damour01} or stochastic background sources \citep{Maggiore,Hobbs_GW,damour05,psh10,Hobbs_GW,Hobbs_Tempo2}. Isotropic stochastic GWB produced by coalescing supermassive binary black hole mergers is the strongest potentially detectable signal of GWs \citep{Hobbs_Tempo2}. Therefore, we use the GWB model to produce synthetic data. The characteristic strain spectrum, $\mathcal{H}_{c}(f)$, for a stochastic GWB can be described by the power-law relation \citep{Hobbs_GW}:
\begin{equation}
\mathcal{H}_{c}(f) = \mathcal{A}_{yr} \left(\frac{f}{f_{1yr}}  \right)^{\zeta}
\label{Amplitude}
\end{equation}
where $f$ is the frequency of GWs, $f_{1yr} \equiv \frac{1}{1yr}$;
$\mathcal{A}_{yr}$ is the dimensionless amplitude of the GWB; and $\zeta$
is a scaling exponent and for almost all expected GWs  is $\zeta
<0$. The corresponding $\zeta$ exponent takes the following values for
different mechanisms:  $\zeta=-\frac{2}{3}$ for coalescing black
hole binaries, $\zeta=-1$ for cosmic strings, and
$\zeta=-\frac{7}{6}$ for primordial GWs from the Big Bang
\citep{Hobbs_GW}. We should mention that the power-law relation obtained in Equation (\ref{Amplitude}) is not unique and there is another framework represented by \citep{Yardley,Sesana}. The
dimensionless amplitude of GWs has been predicted by most authors in
the range of $\mathcal{A}_{yr}\in[10^{-15},10^{-14}]$; however, according
to Refs.~\citep{Yardley,Sesana} the expected range of $\mathcal{A}_{yr}$ for
a stochastic GWB is $\mathcal{A}_{yr}\in [10^{-16} , 3\times10^{-15}]$.

\subsection{Synthetic Data Sets for GWB}
To simulate synthetic series, we use the TEMPO2 software package that carries out the fitting procedure of TOA \citep{tempo2_I}. This package is used to simulate pure timing residuals \citep{Hobbs_Tempo2}.
To simulate the GWB, the "GWbkgrd" plug-in of TEMPO2 will be used \citep{Taylor_GW}.
In the absence of GW signal, we have pure pulsar timing residual represented by  ${\it PTR}_{\rm pure}$, while signal induced by GWB is indicated by ${\it PTR}(t)$.

In order to test the effect of GWs on the $PTR$s, we simulate $100$ timing residuals with 1076 data points that are separated by 13 days with an rms of 100~ns. Then we add the effect of GWB on the simulated pure $PTR$ using different seeds for a given $\mathcal{A}_{yr}$. The chosen accuracy for simulation has been used in other work as a level at which a GWB might be detected \citep{Jenet}; however, it should be noted that only two of the PPTA pulsars (J0437-4715 and J1909-3744) have rms noise of this order (Table \ref{tab:real}).

The GWB introduces two terms for each polarization, one set of which is referred to as the Earth terms. These Earth terms are correlated. However, the other set, referred to as the pulsar terms, has equal amplitude but a long and unknown time delay, so these terms are effectively uncorrelated noise with the same red spectrum as the Earth terms. Our simulations include both the Earth and the pulsar terms.
We simulate 20 pure PTRs for pulsars separated in the sky according to the ephemeris of 20 MSPs observed in the PPTA project (Table \ref{tab:real}). An isotropic GWB induces a particular spatial cross-correlation in $PTR$s leading to a quadrupolar signature ({\it Hellings and Downs} curve) \citep{Hellings,Jenet}. Subsequently, to examine the GWB, we will examine the cross-correlation property of the simulated data.

The upper panel of Fig. \ref{fig:perfect_resi} indicates a typical pure timing residual simulated by TEMPO2 with zero mean uncorrelated series.  We also depict the superposition of pure timing residuals with the GW model introduced in \citep{Hobbs_Tempo2}, in the middle panel of Fig.  \ref{fig:perfect_resi}.
\begin{figure} [t]
\centering
\includegraphics[scale=0.35]{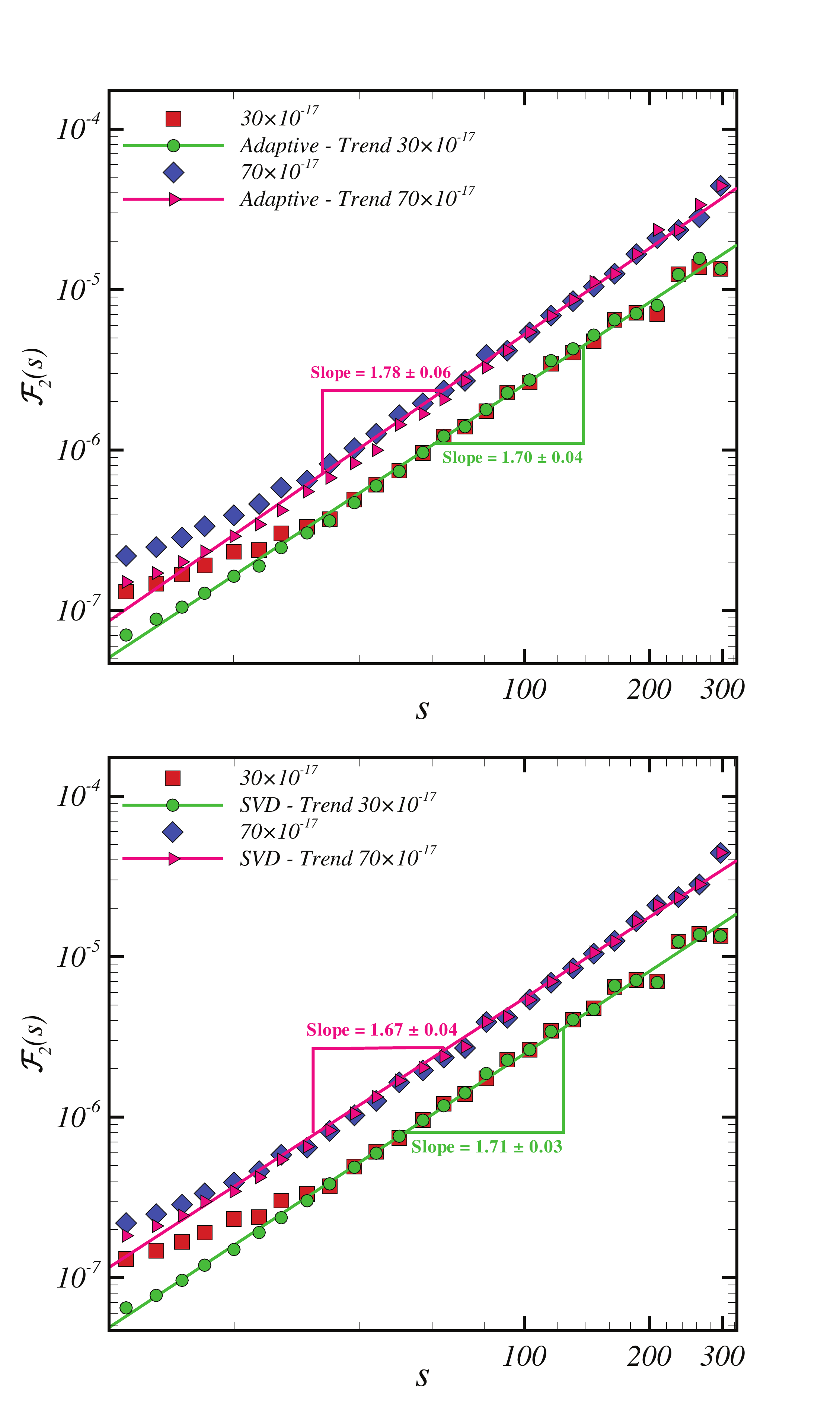}
\caption{ Upper panel: log-log plot of $\mathcal{F}_2(s)$ versus $s$ computed by DMA with $\theta=0.0$ for various simulated {\it PTR}s affected by stochastic GWs when we apply AD as preprocesses. The lower panel is the same as the upper panel but computed by applying SVD as preprocesses. We set $n=100$ in the adaptive method and $p=1$, $d=40$ for the SVD algorithm. In this figure, we take $\zeta=-2/3$. Different values in each plot represent the amplitude of GWs.}
 \label{fspure1}
 \end{figure}

\begin{figure*} [t]
\centering
\includegraphics[scale=0.35]{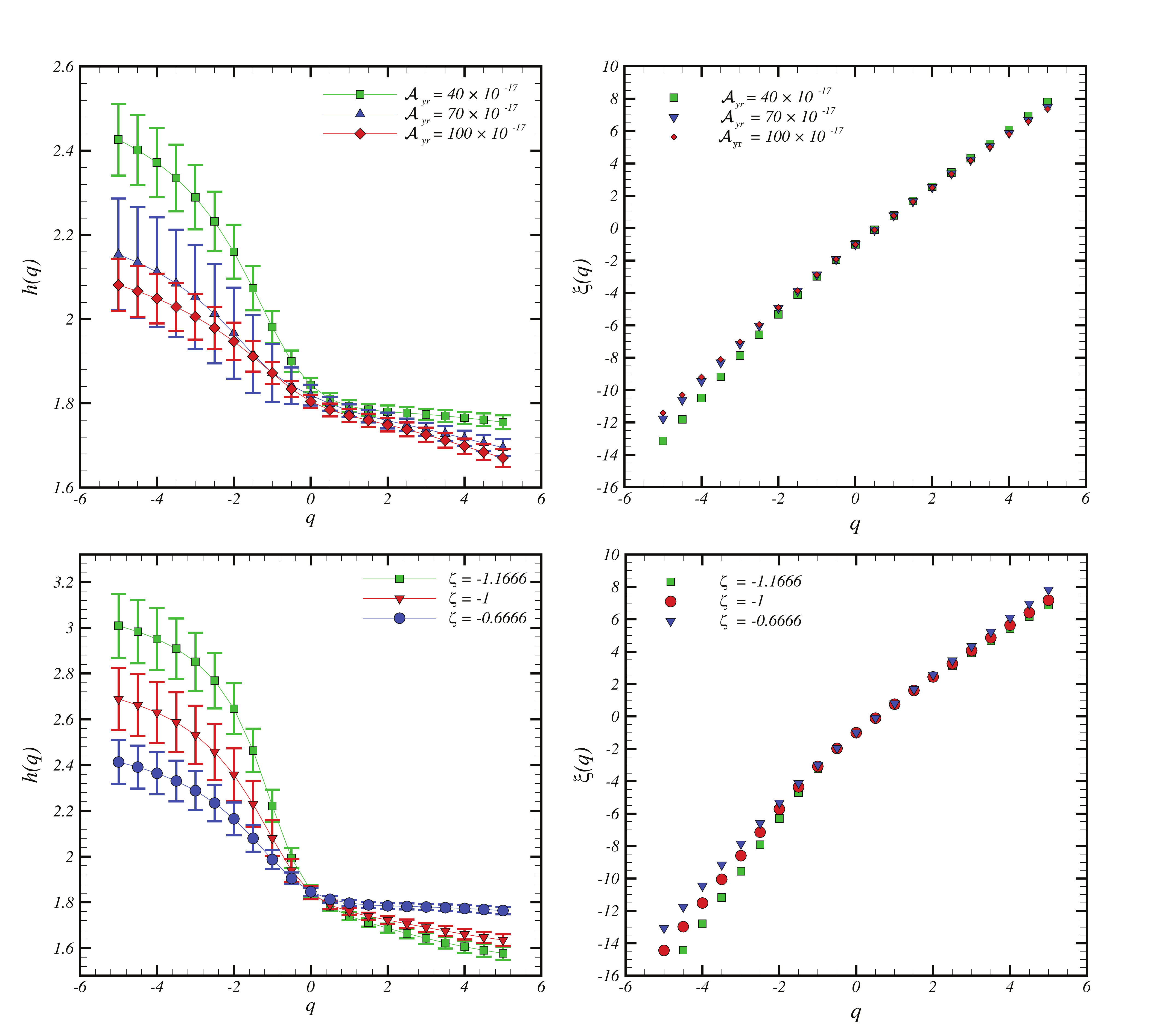}
\caption{The upper left panel indicates the generalized Hurst exponent, $h(q)$, versus $q$ for some $PTR$s induced by stochastic GWs with $\zeta=-2/3$  with various amplitudes calculated  by MF-DMA with $\theta=0.0$. The upper right panel illustrates $\xi(q)$ for the mentioned simulations. The lower panels represent $h(q)$ (left) and $\xi(q)$ (right) for different $\zeta$ with $\mathcal{A}_{yr}=50\times 10^{-17}$.}
\label{hqgw}
\end{figure*}

\subsection{Observed Data}
We use the timing residual data of 20 MSPs observed by the PPTA project at three bandwidths, namely  $10$, $20$, and $50$cm, by implementing the Parkes 64 m radio telescope (PTA) \citep{PPTA}.
The PTA telescope is located in Australia at an altitude of
-33$^{\circ}$ and can observe all of the inner Galaxy.
Due to the higher stability of the short-period MSPs,  the observed pulsars have short periods and are selected from bright ones. Also, these MSPs have narrow pulse widths in order to reduce uncertainties in the corresponding TOA. Finally, isolated wide-binary MSPs have been selected to avoid the effects of the companion star.

The {\it PTR} series for these MSPs as observed datasets are publicly available\footnote{ \url{https://datanet.csiro.au/dap/}}. We have used the TEMPO2 software to extract post-fitted PTRs from timing model data presented by \cite{PPTA}.

The spectralModel\footnote{\url{http://www.atnf.csiro.au/research/pulsar/tempo2}} plug-in is utilized for temporal smoothing and making an equally spaced grid of observed data \citep{coles2011pulsar}. Then, we applied our analysis on post-fitted data.

The names of 20 MSPs with the corresponding rms and total time span are reported in Table \ref{tab:real}. It is worth noting that several phenomena, such as atmospheric delays, vacuum retardation due to observatory motion, Einstein delay, and Shapiro delay, can affect  the TOA \citep{tempo2_II} and they should be dismissed to have a  post-fitted timing residual that is called ${\it PTR}$. The lower panel of Fig. \ref{fig:perfect_resi} illustrates a typical post-fit pulsar timing residual of PSR J0437-4715 observed by the PPTA project \citep{PPTA}. The fitting procedure has been done with the TEMPO2 software.

\begin{figure}
\centering
\includegraphics[scale=0.4]{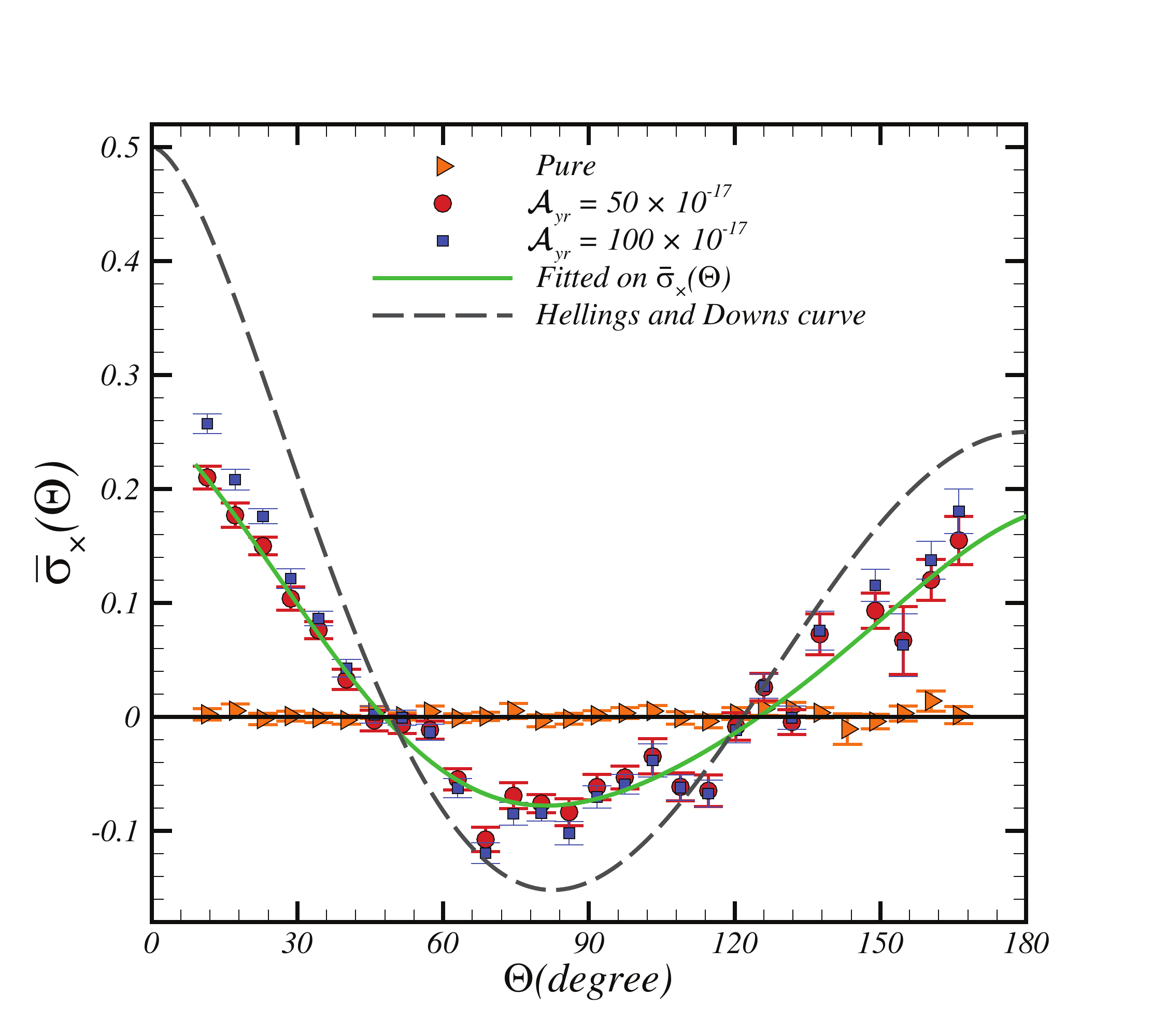}
\caption{The $\bar{\sigma}_{\times}(\Theta)$, versus $\Theta$ for simulated pure (triangles) and induced by stochastic GWB with $\zeta=-2/3$ and $\mathcal{A}_{yr}=50\times 10^{-17}$ (circles), as well as $\mathcal{A}_{yr}=100\times 10^{-17}$ (squares) $PTR$s. By definition, $\bar{\sigma}_{\times}$ is almost insensitive to the value of $\mathcal{A}_{yr}$. The dashed line corresponds to the {\it Hellings and Downs} curve.}
 \label{fig:hdcca}
     \end{figure}

\section{Multifractal analysis of synthetic {\it PTR} series}\label{applyingmethod}
In this section we will evaluate the multifractal nature of synthetic datasets. The capability of our analysis as a detector of gravitational waves and a pipeline for determining the type of GWB will be explained in this section.

\subsection{Multifractal nature of synthetic Data}
At first, we examine the multifractal nature of synthetic ${\it PTR}_{\rm pure}$ and its superposition with simulated GWB. Since, in simulation, our data are regular, therefore we apply common assessment algorithms.
Fig. \ref{fspure} illustrates the fluctuation functions versus scale computed by DMA for $PTR_{\rm pure}$. The results derived by the DFA method are in agreement with the DMA algorithm.
 The average value of the Hurst exponent for all simulated pure $PTR$s is $\langle H\rangle=0.51\pm0.02$ at a $1\sigma$ level of confidence, confirming that $PTR_{\rm pure}$ is an uncorrelated data set \citep{Hobbs_GW}.  Now we superimpose the synthetic  $PTR_{\rm pure}(t)$ with simulated GWB with a given set of free parameters.

 We apply DFA and DMA on simulated $PTR(t)$ for various GWB amplitudes. Fig. \ref{fspure1} illustrates $\mathcal{F}_2(s)$ as a function of $s$ for the simulated series. These results confirm that there is at least one crossover in fluctuation function versus $s$.
We should eliminate the crossover in fluctuation function to determine the generalized Hurst exponent. To this end, we apply either SVD or AD to the datasets, and the clean series are used for further analysis by either the DFA or DMA methods. For SVD, we consider $p=1$ and $d=40$; therefore, the three largest eigenvalues are set to zero, and the new eigenvalues matrix ($\mathbf{\tilde{S}}$), filtered matrix ($\mathbf{\tilde{\Gamma}}$), and cleaned data  ($\widetilde{{\it PTR}}$) are constructed.

Fig. \ref{fspure1} indicates  $\mathcal{F}_2(s)$ computed by the DFA and
DMA algorithms after applying either the SVD or AD method. \cite{Gu.g}
demonstrated that DMA with $\theta=0$ (backward) has the best
performance; therefore, we use the backward DMA method throughout this
paper.  We deduce that applying an SVD preprocess can efficiently
remove the crossover, and we are able to assign a scaling exponent
for fluctuation function versus $s$. The situation for  AD
preprocessing is somehow different, but it is consistent with the SVD
results. The generalized Hurst exponent and $\xi$ versus $q$ for
three types of $PTR$s superimposed by different values of GWB
amplitudes are depicted in Fig. \ref{hqgw}. The upper panels of Fig.
\ref{hqgw} illustrate the $h(q)$ and $\xi(q)$ for synthetic $PTR$s
affected by GWB with different amplitudes with the same $\zeta$. As we
expect, the value of $h(q=2)$ that is related to $\zeta$ for all
samples is almost same. The lower panel shows $h(q)$ and
$\xi(q)$ for simulated $PTR$s with different $\zeta$.

\subsection{Irregular MF-DXA of simulated $PTR$s}
The quadrupolar signature of the spatial cross-correlation function of $PTR$s is considered as a particular measure for detecting the imprint of the GWB \citep{Taylor:2016gpq}. Previously, the {\it Hellings and Downs} curve has been examined for detection of stochastic GWB  \citep{Hellings,Jenet,Taylor:2016gpq}.

Implementation of Irregular MF-DXA on $PTR$s provides a reliable cross-correlation exponent and coefficient in the presence of nuisance trends and noises. Irregular MF-DXA is indeed a crucial part of our pipeline for searching the significance of GWB. Here, due to the regularity of the simulated data, we consider the usual MF-DXA. To show the validity of this idea, we simulate 20 pure $PTR$s for pulsars separated in the sky according to the ephemeris of 20 MSPs observed in the PPTA project given in Table \ref{tab:real}. Then, we add the effect of GWB to each pure $PTR$.

In Fig. \ref{fig:hdcca}, we show $\bar{\sigma}_{\times}(\Theta)$ for simulated $PTR$s. Here we have simulated 50 realizations for 20 pulsars. The points plotted in Fig. \ref{fig:hdcca} are the average of these 50 realizations.
As indicated in this figure, when synthetic $PTR$s are affected by GWB with $\mathcal{A}_{yr}=50\times 10^{-17}$ and $\zeta=-2/3$, we can recognize a quadrupolar feature in $\bar{\sigma}_{\times}(\Theta)$ which is a benchmark for existing GWB. This behavior is similar to the {\it Hellings and Downs} curve indicated by the dashed line in Fig. \ref{fig:hdcca}. One of the advantages of this new measure is that, when undesired parts exist in the observed series, we are able to infer the contribution of the GWB signal robustly.  Eqs. (\ref{eq:newsigma0}) and (\ref{eq:newsigma}) also confirm that $\bar{\sigma}_{\times}$ is almost insensitive to the value of $\mathcal{A}_{yr}$.
To make a more conservative pipeline for assessing the GWB signal, it is necessary to compute the cross-correlation coefficient, $\bar{\sigma}_{\times}(\Theta)$, in addition to the usual spatial cross-correlation function known as the {\it Hellings and Downs} curve. After obtaining  the feature, we carry out the rest part of the MF-DMA analysis to determine the type and amplitude of GWB signal.

\begin{figure*} [t]
\centering
\includegraphics[scale=0.35]{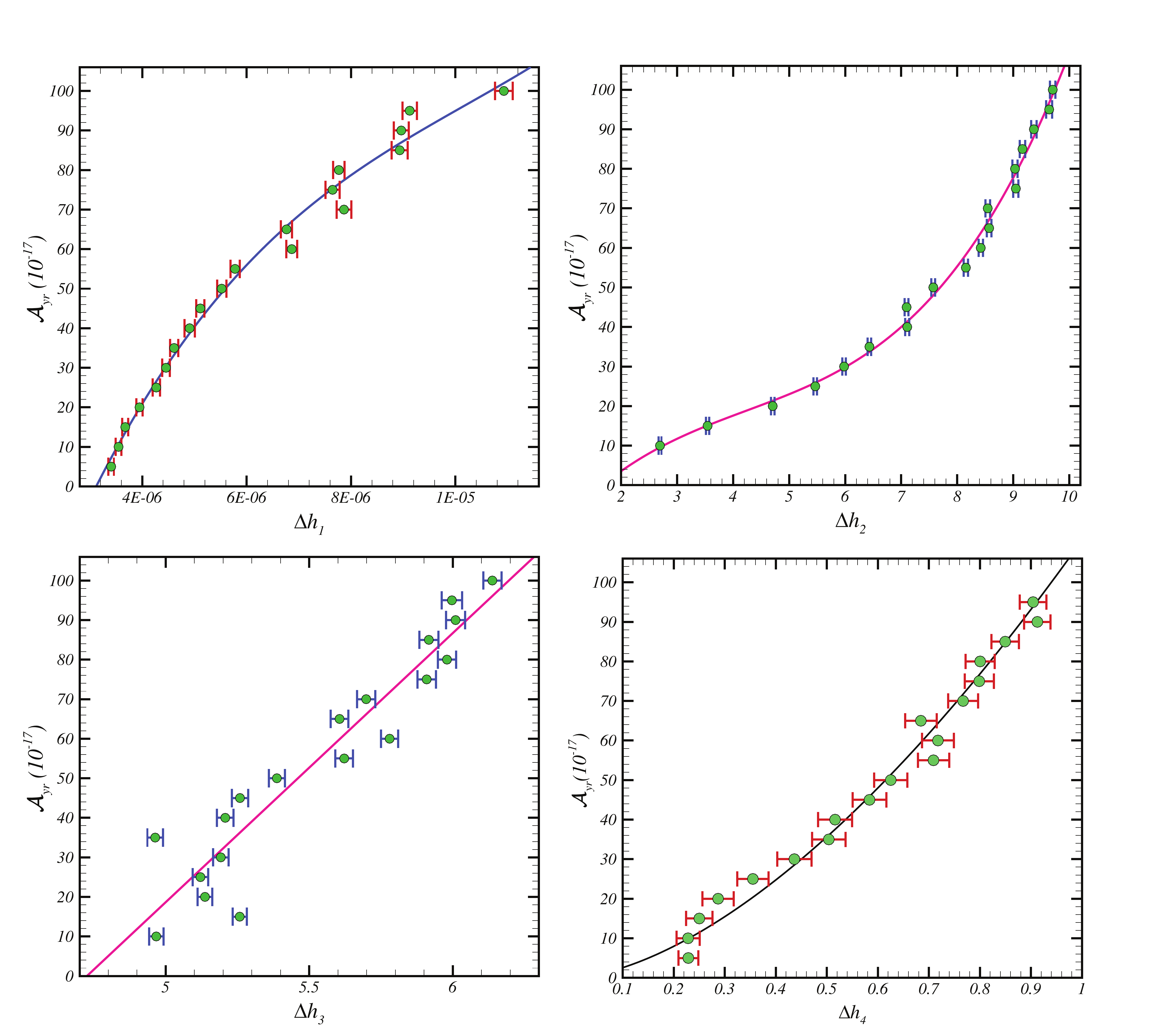}
\caption{Value of $\mathcal{A}_{yr}$ determined by four strategies introduced in this paper only for $\zeta=-2/3$ and rms=100~ns. The solid lines are typical fitting functions.}
 \label{fig:strategy}
     \end{figure*}

\begin{figure*} [t]
\centering
\includegraphics[scale=0.6]{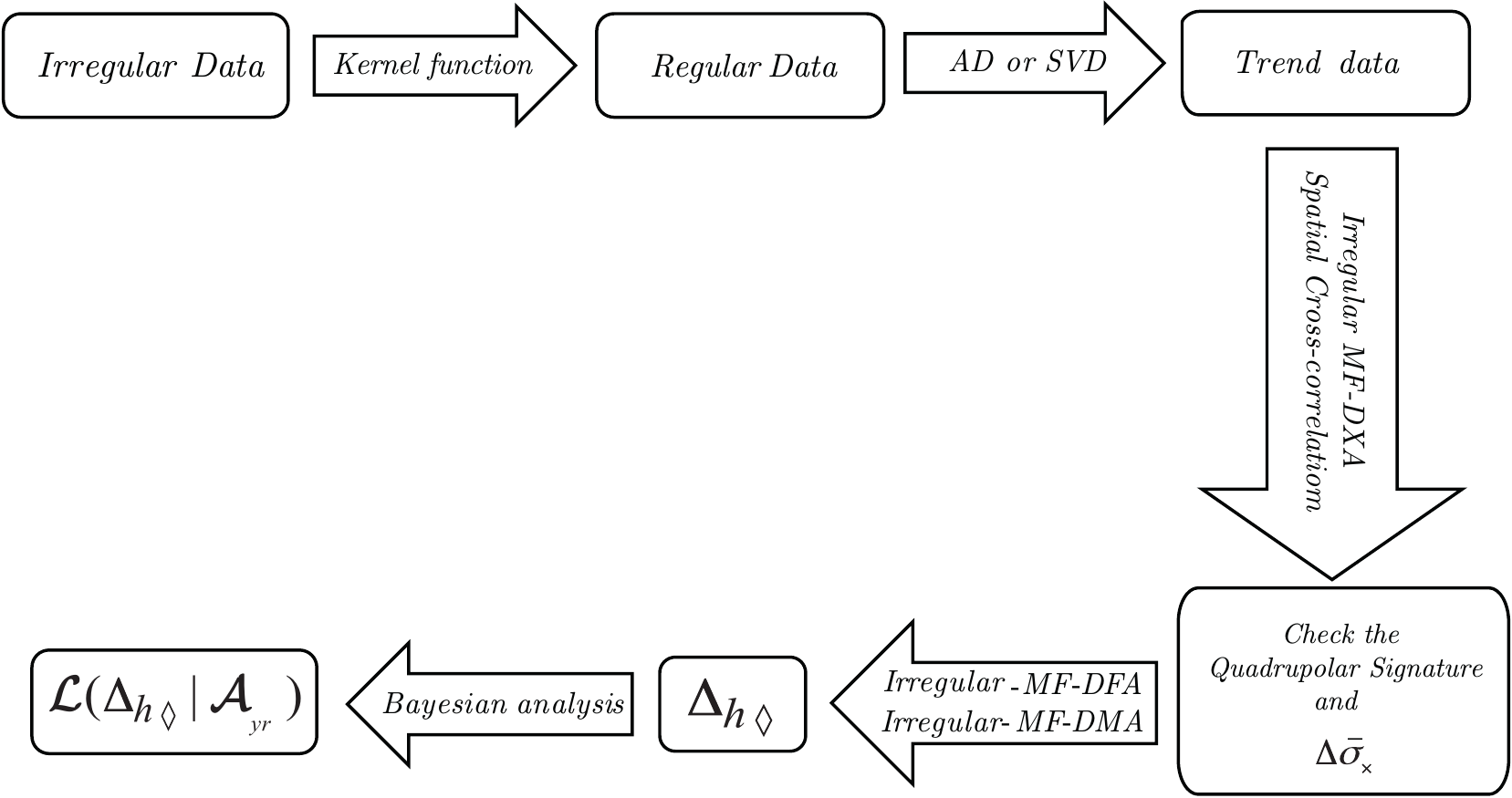}
\caption{Schematic representation of our pipeline for searching the footprint of GWB in the context of multifractal analysis of irregular $PTR$s.}
 \label{fig:pip}
\end{figure*}

\subsection{Strategies for Searching GWs}\label{startegy}
According to the results presented in the previous sections, the randomness of pure $PTR$s exhibits that deviations from uncorrelated behavior can be considered as additional features presented in the recorded data. Unfortunately, the observed {\it PTR}s may include intrinsic fractal noise, interstellar plasma, uncertainties in the Earth's motion, master clocks, and receiver signals. It has been demonstrated that the noise from some of these sources is wavelength dependent and  has spatial correlation, either monopole or dipole in nature. Subsequently, relying on multifractal analysis modified by preprocessing algorithms such as the AD or SVD methods of individual {\it PTR}s probably gives rise spurious results in the framework of GW searching. To get rid of the effect of undesired components, we rely on the quadrupole structure of the GWB and carry out the irregular MF-DXA  approach.

Therefore, we begin with  Irregular-MF-DXA on all available $PTR$s distributed over all directions and then compute $\bar{\sigma}_{\times}(\Theta)$ as a function of separation angle, $\Theta$. The existence of a feature similar to Fig. \ref{fig:hdcca} in observed $PTR$s would imply detection of a GWB. Note that Fig. \ref{fig:hdcca} is the average of 50 realizations. One observation with these parameters would have error bars almost 7 times larger, so the GWB would be detected but the significance would be much less. Thereafter,  we will turn to the multifractal behavior of the  $PTR$ series to determine the type and amplitude of the GWB.
In order to determine the type of stochastic GWB with a strain spectrum modeled by Eq. (\ref{Amplitude}), after preprocessing to remove noise and foreground, we apply multifractal methods to compute a reliable Hurst exponent. This exponent is related to the power-spectrum exponent. Finally, the best-fit value of $\zeta$ and its associated error bar are determined \citep{Hobbs_Tempo2}. However, there are many complications in the real data sets, making the inference procedure less straightforward to assess GWs. We therefore introduce four criteria as follows:

$I)$ According to Eqs. (\ref{eq: fluctuation function}) and (\ref{eq:g}), the intercept of fluctuation function for  {\it PTR}s contains the intensity of  superimposed GWs. Therefore, after recognizing a   quadrupolar signature in analyzing pairs of {\it PTR}s, the following quantity is able to indicate the intensity of GWB:
$\Delta h_1(\mathcal{A}_{yr},\zeta)\equiv \sum_{q=q_{min}}^{q=q_{max}} | \mathcal{G}_{h(q)}(\mathcal{A}_{yr},\zeta)-\mathcal{G}_{h(q)}(\mathcal{A}_{yr}=0)|$.  In practice, we find a robust mathematical relation between
$\Delta h_1(\mathcal{A}_{yr},\zeta)$ and $\mathcal{A}_{yr}$ for any given $\zeta$ (or, equivalently, $H$) and rms of white noise, as follows. We do many simulations for a given value of $\zeta$ with different $\mathcal{A}_{yr}$ values. Then, we apply either SVD or AD to make clean data. The clean data are used for further analysis. According to our simulation for $\zeta=-2/3$ and rms=100~ns,  the mathematical relation between $\mathcal{A}_{yr}$ and $\Delta h_1$  in the range of $\mathcal{A}_{yr}\in[10^{-17},10^{-15}]$  reads as:
\begin{eqnarray}\label{measure11}
 \left(\frac{\mathcal{A}_{yr}}{10^{-17}}\right)=  a \Delta h_1^2 + b\Delta h_1 + c
\end{eqnarray}
where $a =  (-1.15\pm0.40)\times 10^{12}$, $b = (2.84\pm0.54)\times 10^{7}$ and $ c = -74.45\pm16.88$. This fitting function is not unique, and here we select one with a high goodness of fit before going further. Also, for any other rms dictated by experiment, the above analysis should be repeated again to find the corresponding fitting function.

$II)$ For pure {\it PTR}s, we found that the Hurst exponent is almost $0.5$, while there will be deviations in the generalized Hurst exponent for $PTR$ signals affected by GWs (Eq. (\ref{Amplitude})) for a given amplitude $\mathcal{A}_{yr}$, and $\zeta$.  Therefore, another powerful measure to quantify the intensity of the GWB would be  $\Delta h_2(\mathcal{A}_{yr},\zeta)\equiv \sum_{q=q_{min}}^{q_{max}} |h(q;\mathcal{A}_{yr},\zeta)-h_{\rm shuf}(q;\mathcal{A}_{yr},\zeta)|$. Where $h_{\rm shuf}(q;\mathcal{A}_{yr},\zeta)$ is for completely randomized $PTR$ and "shuf" refers to shuffled.   In practice, we find a robust mathematical relation between
$\Delta h_2(\mathcal{A}_{yr},\zeta)$ and $\mathcal{A}_{yr}$ for any given $\zeta$ (or, equivalently, $H$) and rms of white noise. 
The corresponding shuffled series are produced using original series. Now by calculating the generalized Hurst exponent for original and shuffled data, one can compute $\Delta h_2$.
We find that the following function is a good fit to our simulations for $\mathcal{A}_{yr}$ in the range of $\mathcal{A}_{yr}\in[10^{-17},10^{-15}]$ versus $\Delta h_2$ for  $\zeta=-2/3$ and rms=100~ns:
\begin{eqnarray}\label{measure2}
%\mathcal{A}_{yr} &=& a\exp(b\Delta h_2) + c\exp(d\Delta h_2)
\left(\frac{\mathcal{A}_{yr}}{10^{-17}}\right)&=&a\Delta h_2^3+b\Delta h_2^2+c\Delta h_2
\end{eqnarray}
where $a=0.19\pm0.06$, $b=-1.57\pm0.92$, and $c= 7.40\pm3.30$. This fitting function is not unique, and here we select a high goodness of fit. Before going further, it is worth noting that  the whitened noise generation is serious in many simulations. An optimal algorithm to evaluate noise quality in many simulations, especially in data generation by the TEMPO2 software, can be carried out  by the shuffling procedure explained here. Subsequently, our proposal in this regard can be straightforwardly implemented as a new plug-in.

$III)$ Since GWs may induce non-Gaussianity in $PTR$, it is
interesting to  take into account $\Delta
h_3(\mathcal{A}_{yr},\zeta)\equiv \sum_{q=q_{min}}^{q_{max}} |h(q;\mathcal{A}_{yr},\zeta)-h_{\rm
sur}(q,\mathcal{A}_{yr},\zeta)|$. In the mentioned criterion, $h_{\rm sur}(q;\mathcal{A}_{yr},\zeta)$ is the generalized
Hurst exponents computed for Gaussian datasets with the same
correlation function as the original series. Here "$\rm sur$"
represents surrogated data or phase-randomized
surrogated series, including the multiplication of Fourier-transform
data by a random phase with a uniform distribution function \citep{prich94}. We
simulated the $PTR$ accompanying the GWB with different amplitudes, and
the following fitting function is determined for $\mathcal{A}_{yr}$ in the same range as above versus $\Delta h_3$ for $\zeta=-2/3$ and rms=100~ns:
\begin{eqnarray}\label{measure3}
\left(\frac{\mathcal{A}_{yr}}{10^{-17}}\right)&=& a\Delta h_3 + b
\end{eqnarray}
where $a = 68.03\pm11.73$ and $b=-321.50\pm65.10$.

$IV)$ The width of the singularity spectrum, which quantifies the nature of multifractality, is  another benchmark for determining the amplitude of GWs superimposed on the $PTR$s. This measure is defined by $\Delta h_4(\mathcal{A}_{yr},\zeta) \equiv |\Delta \alpha(\mathcal{A}_{yr},\zeta)- \Delta \alpha(\mathcal{A}_{yr}=0)|$. According to our simulations,  we find:
\begin{eqnarray}\label{measure4}
\left(\frac{\mathcal{A}_{yr}}{10^{-17}}\right) &=& a\Delta h_4^b+c %\exp(b\Delta h_3) + c\exp(d\Delta h_3),
\end{eqnarray}
for $\zeta=-2/3$ and rms=100~ns in the range of $\mathcal{A}_{yr}\in[10^{-17},10^{-15}]$. Here $a=106.30\pm7.80$, $b=1.62\pm0.42$, and $c= 1.52\pm9.74$.\\ %and $d=-4.81$.\\

Let us summarize our strategy based on the above criteria for searching GWs in observation. As explained in section 2, in the case of the proper value of signal-to-noise (S/N) for each observed {\it PTR}, we remove all known contributions  from foreground contamination. Therefore, we make regular series according to methods explained in subsection 2.3. Now we are ready to apply either AD or SVD method to extract the dominant part of the signal (the trend part) from the noise. Then, we apply the MF-DXA method to compute $\bar{\sigma}_{\times}$, and  we compute the spatial cross-correlation to identify the probable quadrupolar signature. In the case of finding the mentioned signature, we go through the detection of GWs. Otherwise, we can only carry out the upper-limit approach. We also apply irregular MF-DXA on the proper part of the series for all available pairs of observed $PTR$s to examine the temporal part of the cross-correlation function and deduce the temporal scaling exponent. In the case of the homogeneous and isotropic source of the GWB, $h_{\times}$ is independent from the angular separation of $PTR$s, while for anisotropic or different single sources of GWs, the scaling exponent of the temporal part of the cross-correlation gets various values for different pairs. Utilizing  either irregular MF-DFA or irregular MF-DMA  on cleaned data leads to computing $h(q)$. The best-fit value of  $\zeta$ is then determined by using the power-spectrum exponent.  Following the benchmarks, we compute $\Delta h_1$, $\Delta h_2$, $\Delta h_3$, and $\Delta h_4$ for the observed $PTR$s. The GWB amplitude can be conservatively read  from the corresponding plots, as indicated in Fig. \ref{fig:strategy} or stated by Eqs (\ref{measure11}), (\ref{measure2}), (\ref{measure3}) and (\ref{measure4}). It is worth noting that the functional form of $\Delta h$ should be determined for each value of $\zeta$ and given rms of white noise associated with observed data.  Finally, we are able to compute the upper limit  on $\mathcal{A}_{yr}$ using posterior analysis (see section \ref{applyingmethod2}).  Fig. \ref{fig:pip} is a schematic representation of the pipeline.

Here we emphasize some important considerations for dealing with observed $PTR$s. First of all, we define a relative difference between the scaling exponent computed for the observed $PTR$s and that computed for the $PTR$s without GWB to reduce the contribution of noise and trends.
Finally, in our approach, the level of noise is almost no longer serious when we focus on the scaling exponent.

\begin{figure} [t]
\centering
\includegraphics[scale=0.35]{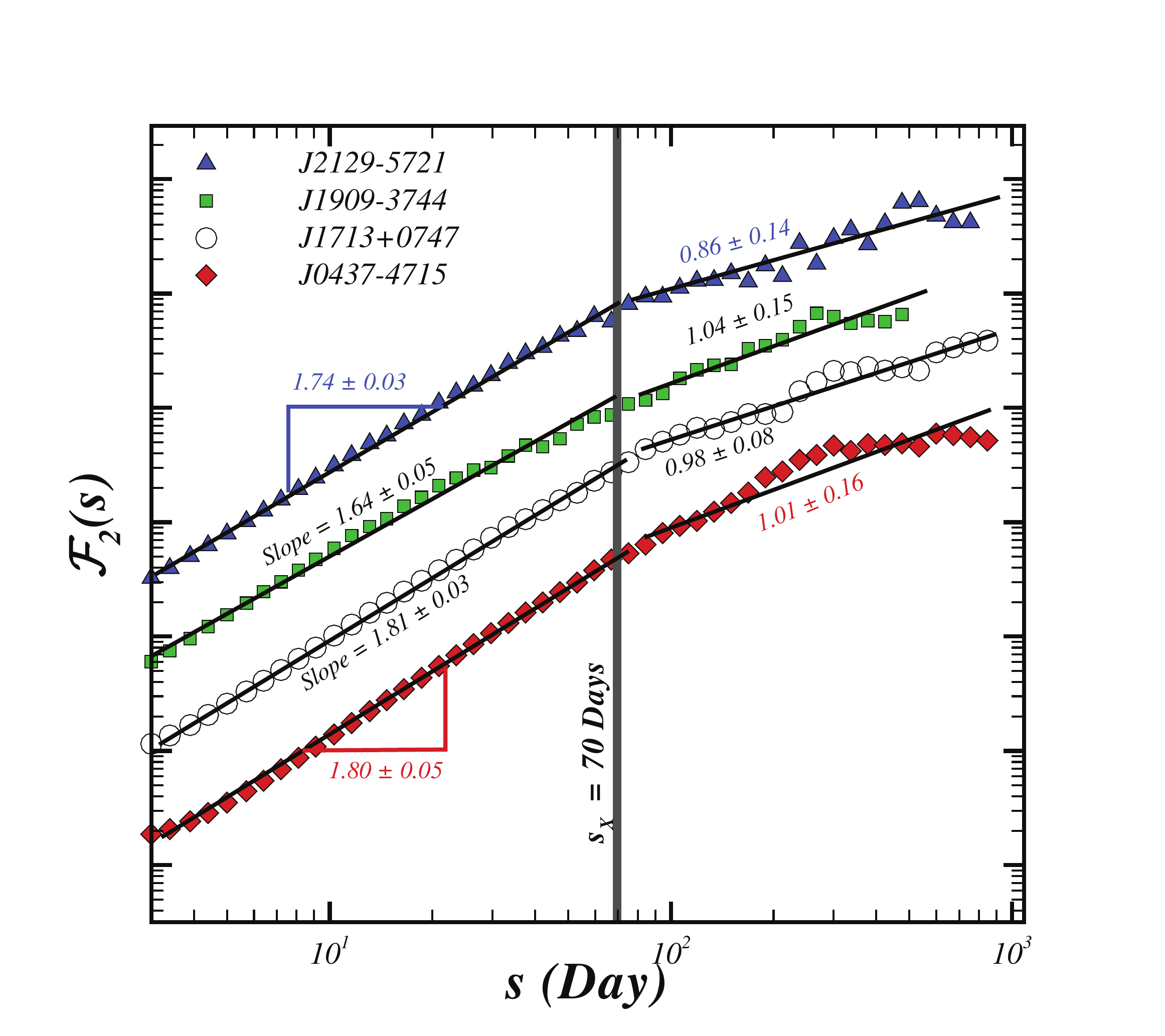}
\caption{Log-log plot of $\mathcal{F}_2(s)$ versus $s$ computed according to backward DMA, namely $\theta=0.0$, for various observed datasets. To make more sense, we shifted $\mathcal{F}_2$ vertically for different amplitudes}
 \label{fig:different_type}
 \end{figure}

\begin{figure} [t]
\includegraphics[scale=0.50]{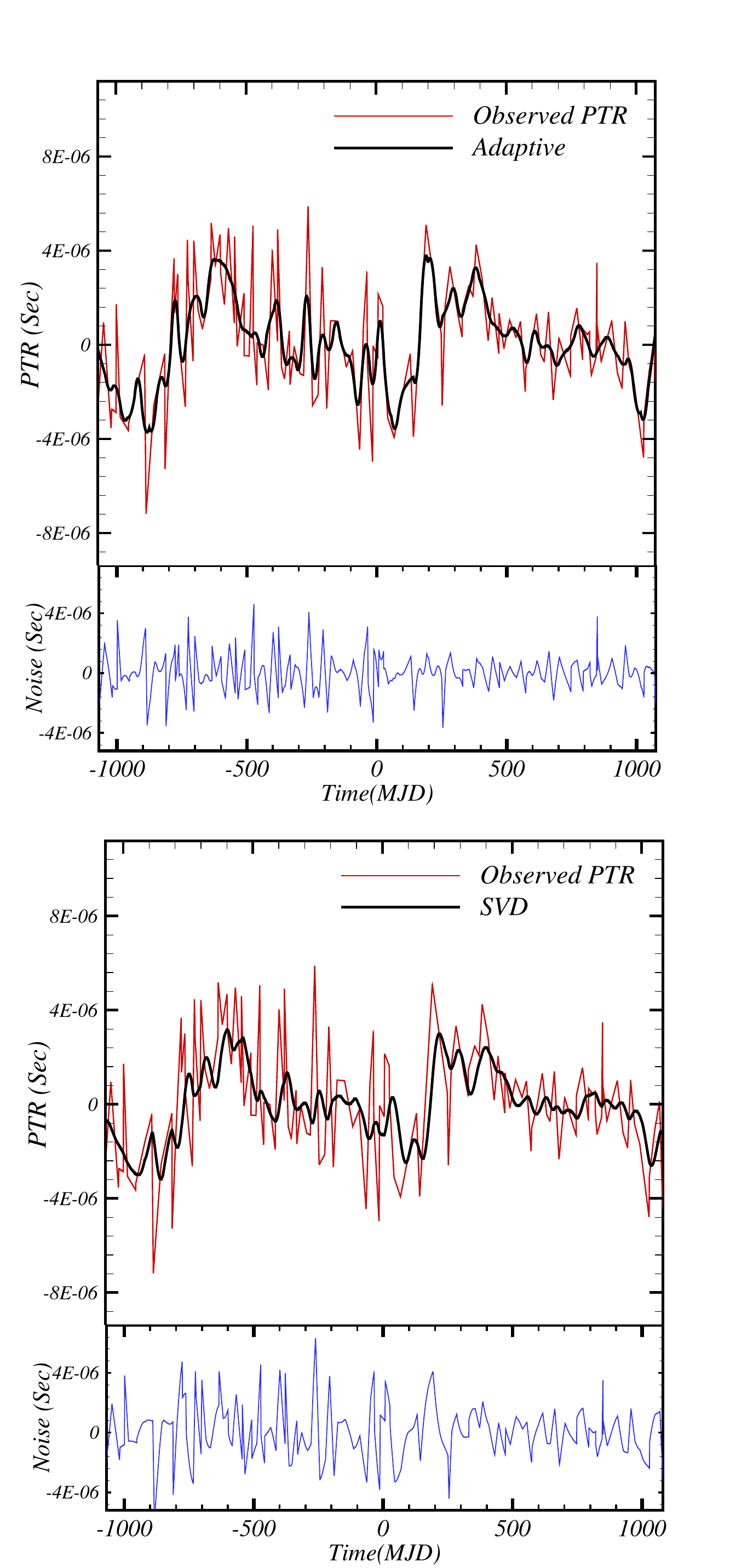}
\caption{Implementing of AD (upper panel) and SVD
(lower panel) on the $PTR$ of PSR J1603-7202. In each panel, the top plot
corresponds to the observed data (red line) and trend (black line), while the bottom represents the residual data
corresponding to clean data.}
 \label{fig:adaptiveapp}
 \end{figure}

\begin{figure} [t]
\centering
\includegraphics[scale=0.35]{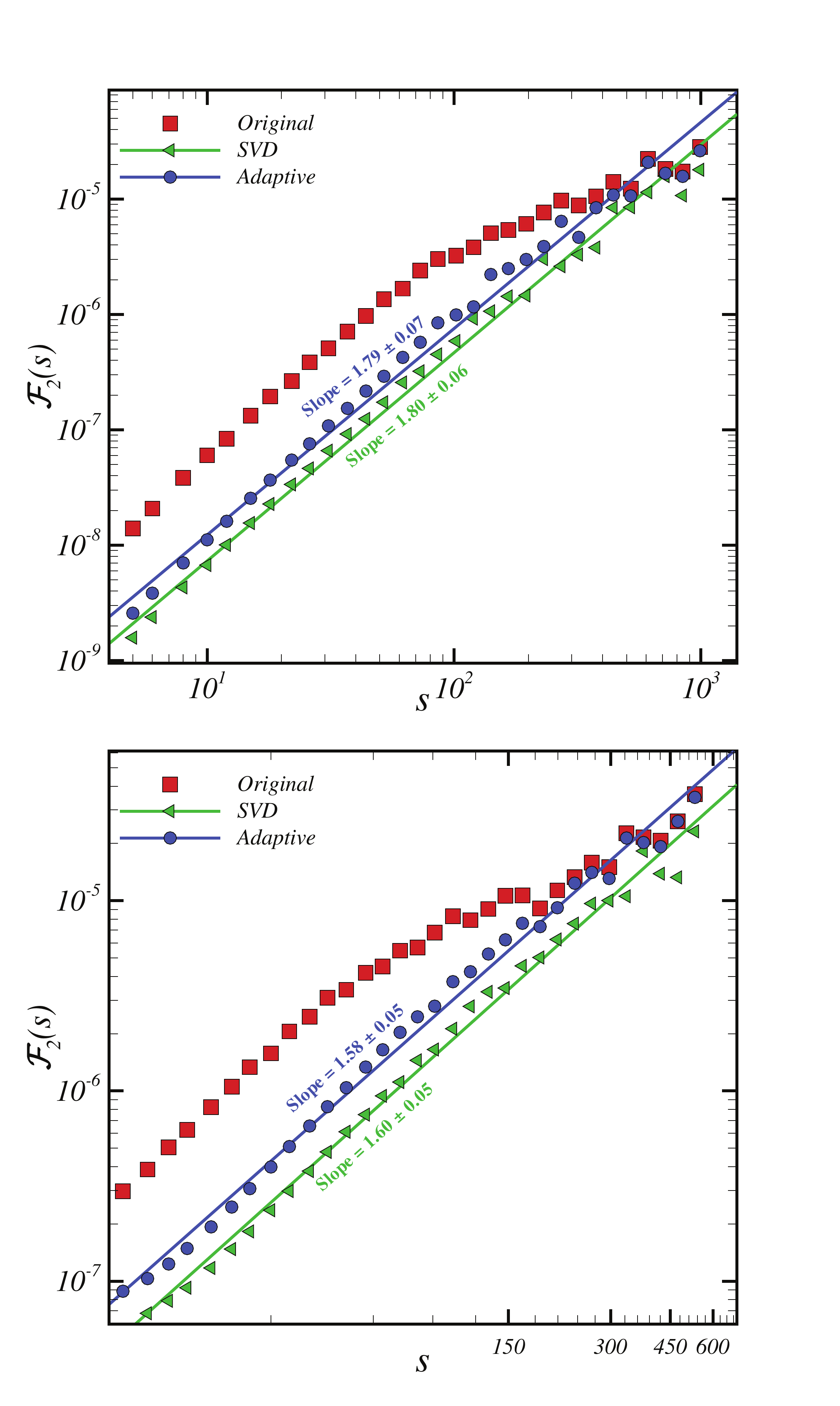}
\caption{Log-Log plot of fluctuation function $\mathcal{F}_2(s)$ as a function of $s$ when we apply AD and SVD as preprocesses on PSR J1857+0943. The upper panel is for DFA, while the lower panel is for backward DMA.}
 \label{fig:dfa_cross}
 \end{figure}

\begin{figure} [t]
\centering
\includegraphics[scale=0.35]{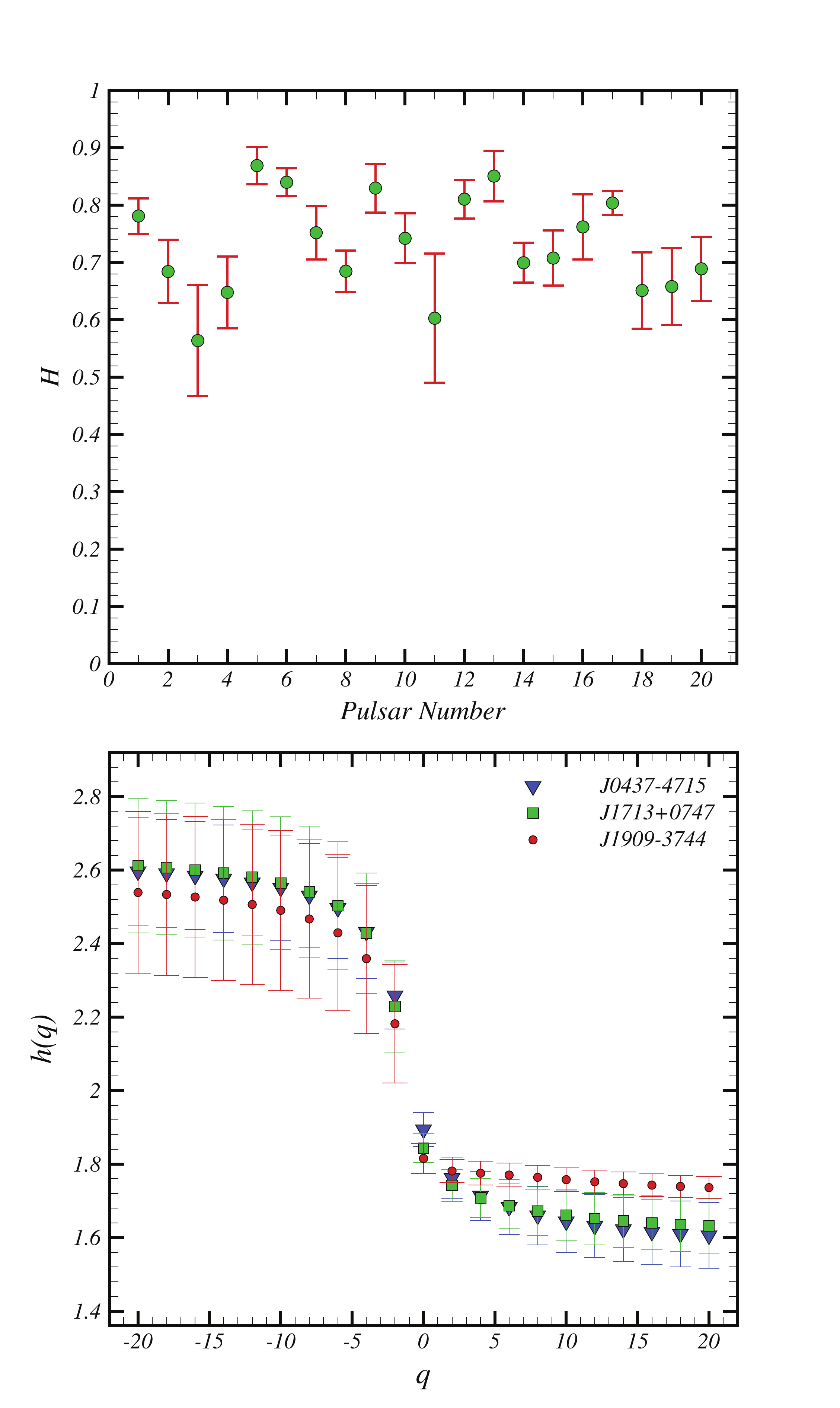}
\caption{Upper panel: Hurst exponent of timing residuals of 20 MSPs observed by PPTA. Lower panel: generalized Hurst exponent $h(q)$ versus $q$ by the SVD-MF-DMA method with $\theta = 0.0$ for some observed timing residuals.}
 \label{fig:real_h(2)}
 \end{figure}

\begin{figure} [t]
\centering
\includegraphics[scale=0.36]{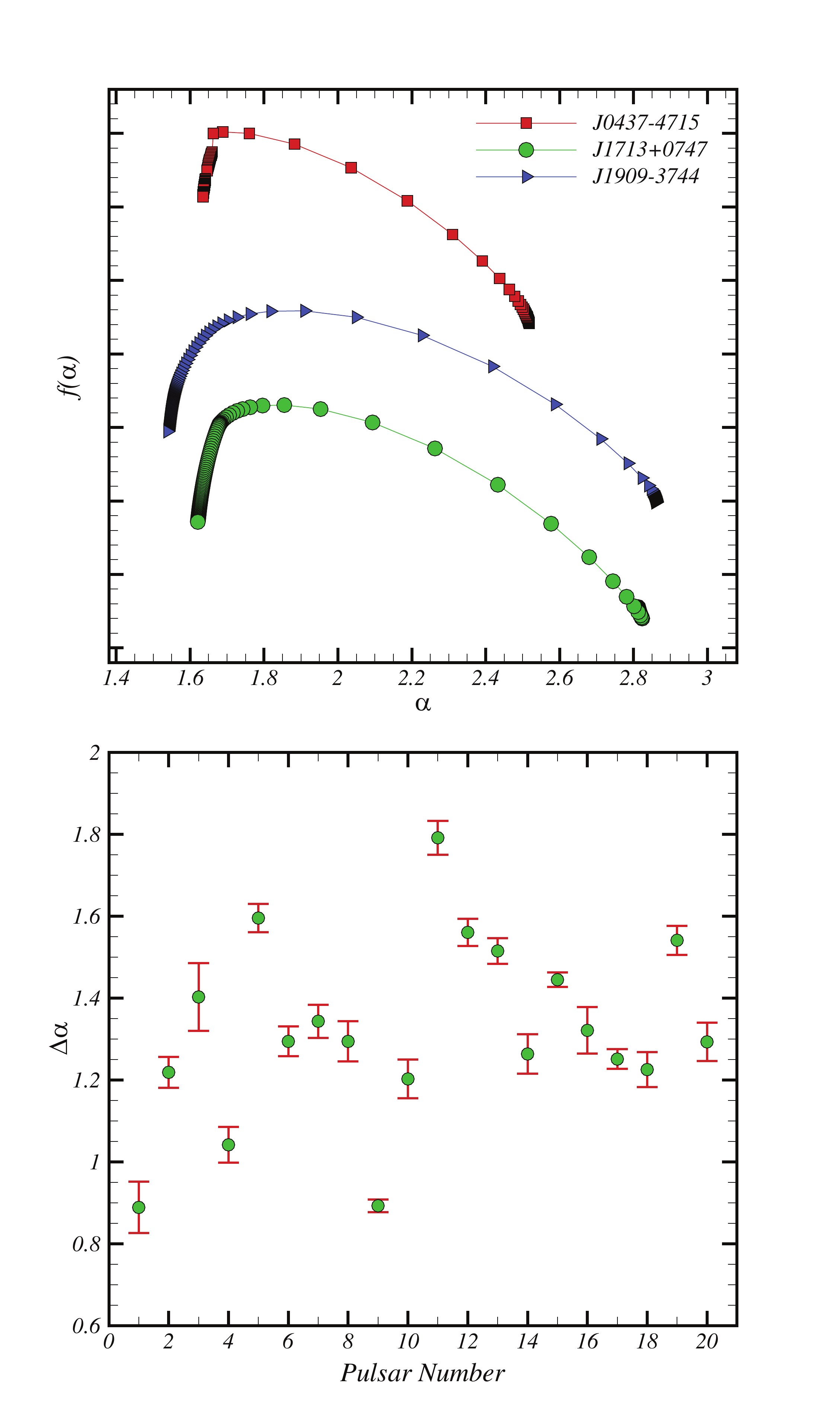}
\caption{The upper panel shows the singularity spectrum $f(\alpha)$ versus $\alpha$ for some observed timing residuals. To make it more obvious, we shifted $f(\alpha)$ vertically for different series. The lower panel indicates the width of the singularity spectrum, which is a measure for quantifying the multifractal nature of 20 MSPs observed by PPTA. }
 \label{fig:f_alpha}
 \end{figure}

\section{Implementation of multifractal methods on observed {\it PTR} data }\label{applyingmethod2}
 Here we use the MF-DFA and MF-DMA methods modified by either AD or SVD detrending procedures  to  examine the multifractal and complexity  behavior of observed pulsar timing residuals.
\subsection{Implementation on Observed Data}
As discussed in subsection 2.3,  observed $PTR$s datasets are in the form of irregularly sampled series, and here we use the spectralModel plug-in for the temporal smoothing  algorithm  to construct equidistant regular series for further analysis \citep{coles2011pulsar}. The size of the current observed data is not large enough to use the irregular version of MF-DFA and MF-DMA introduced by Eqs. (\ref{fluc2new}) and (\ref{fluc1new}).

Fig. \ref{fig:different_type} illustrates the MF-DMA results for various observed {\it PTR}s. These results confirm that there is a crossover in fluctuation functions versus $s$, corresponding to $s_{\times}\sim70$ days.
For the scaling exponent for $s<s_{\times}$, we have $h(q=2)\in[1.03,1.82]$, demonstrating that datasets have a nonstationary nature, while  for $s>s_{\times}$, we find $h(q=2)\in[0.07,1.55]$.

In order to get rid of these crossovers and have a scaling behavior in fluctuation functions, we apply either AD or SVD separately on modified observed datasets. Then, the cleaned data will be used as input for the MF-DFA and MF-DMA algorithms. Fig. \ref{fig:adaptiveapp} illustrates a typical observed $PTR$ (red line) and the trend (black line) determined by AD (upper panel) and SVD (lower panel). The corresponding residual between the observed data and trend is indicated in the bottom of this figure. Fig. \ref{fig:dfa_cross} represents the fluctuation functions computed for a typical observed $PTR$ by DFA and DMA applied on cleaned data provided by AD and SVD separately. The slope of the fluctuation functions for $q=2$ in reliable scales is $h(q=2)\in [1.56,1.87]$, demonstrating that all underlying series are categorized in the nonstationary class. The corresponding Hurst exponent, $H=h(q=2)-1$, belongs to $H\in[0.56,0.87]$. The value of the Hurst exponents for all observed  {\it PTR}s at the $68\%$ level of confidence is depicted in Fig. \ref{fig:real_h(2)}. This result confirms that the dominant part of  observed {\it PTR}s belongs to the long-range correlated signal. The lower panel of Fig. \ref{fig:real_h(2)} shows the $q$-dependency of the generalized Hurst exponent after applying SVD on observed data and determined by MF-DMA. The results for MF-DFA are consistent with those  determined by  MF-DMA. Since $h$  depends on $q$, we conclude that all observed {\it PTR}s are multifractal.  Singularity spectra of some observed {\it PTR}s are plotted in the upper panel of Fig. \ref{fig:f_alpha}. The strength of the multifractal nature of {\it PTR}s is determined by the width of the singularity spectrum, $\Delta \alpha=\alpha_{\rm max}-\alpha_{\rm min}$. This value for observed data is reported in Table \ref{tab:real} and is also shown in the lower panel of Fig. \ref{fig:f_alpha}. The range of the mentioned singularity spectra is $\Delta \alpha\in [0.89,1.79]$. Other relevant exponents are reported in Table \ref{tab:real}.

An interesting question is, what are the sources of multifractality of observed {\it PTR}s? As explained in more detail by \cite{Kantelhardt},  in principle, different correlation functions at small and large fluctuations can be considered as a source of multifractality. In addition,  heavy-tailed probability distribution contributes to the multifractal behavior.
In order to distinguish the two mentioned  types of multifractality, we follow the method introduced in \citep{Kantelhardt}. By shuffling the series, the scaling behavior of the ratio of fluctuation functions, $\mathcal{F}_q(s)/\mathcal{F}_q^{\rm shuf}(s)$, is represented as:
\begin{equation}
\frac{\mathcal{F}_q(s)}{\mathcal{F}_q^{\rm{shuf}}(s)} \sim s^{h(q)-h_{\rm {shuf}}(q)}
\label{shuf}
\end{equation}
where $h_{\rm{shuf}}(q)$ is the generalized Hurst exponent for shuffled data.
The case of $h_{\rm {cor}}(q)\equiv h(q)-h_{\rm {shuf}}(q)=0$ refers to multifractality sourced by the distribution function. In this case, we can compute $h_{\rm PDF}(q)\equiv h(q)-h_{\rm sur}(q)$. If both $h_{\rm {cor}}(q)$ and $h_{\rm{PDF}}(q)$ depend on $q$, both sources are playing roles in the multifractality of the data. In our samples, all  {\it PTR}s have $h_{\rm {shuf}}(q)=0.50$ at a $1\sigma$ confidence interval, confirming that the correlation in datasets  is almost the main source of multifractality.
This property is a universal feature of all observed {\it PTR}s investigated in this paper.

\begin{figure} [t]
\centering
\includegraphics[scale=0.48]{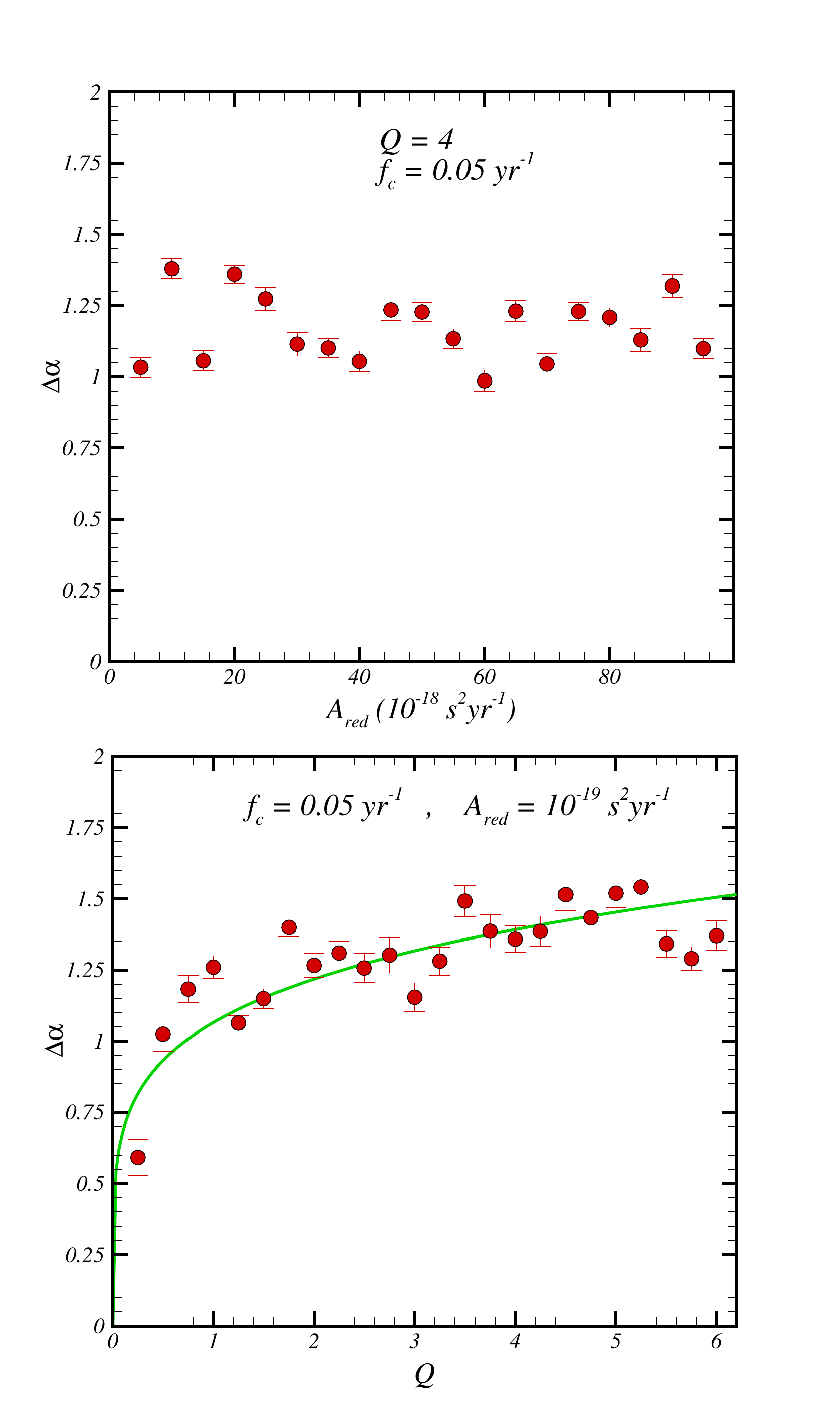}
\caption{The upper panel shows the width of the singularity spectrum as a function of the red-noise power-spectrum amplitude. The lower panel indicates the width of the singularity spectrum versus $Q$. }
 \label{fig:redmodel}
 \end{figure}

The multifractality responsible for observed {\it PTR}s can also be examined by our method.  To this end, we have used different models for the noise component according to the SimRedNoise plug-in of TEMPO2 and applied the MF-DMA method on those series. The upper panel of Fig. \ref{fig:redmodel} indicates that the width of the singularity spectrum computed by the MF-DMA method is almost independent of the amplitude of the red-noise model. The lower panel illustrates the dependency of $\Delta \alpha$ on the exponent of the red-noise power spectrum considered as the $P_{red}(f)=A_{red}(1+f^2/f_c^2)^{-Q/2}$ model, where $A_{red}$, $f_c$, and $Q$ are the amplitude of the power spectrum, corner frequency, and power-law index, respectively \citep{hobbs2006pulsar,Arch16}. In this equation, $Q=0$ corresponds to white noise, and $Q=2,4$, and $6$ are related to phase noise, frequency noise and spin-down noise. Subsequently, we can deduce that the red noise can be responsible for multifractality of observed PTRs as well as GWs. Therefore, a part of our reported multifractality is related to red noise.

\begin{figure} [t]
\centering
\includegraphics[scale=0.35]{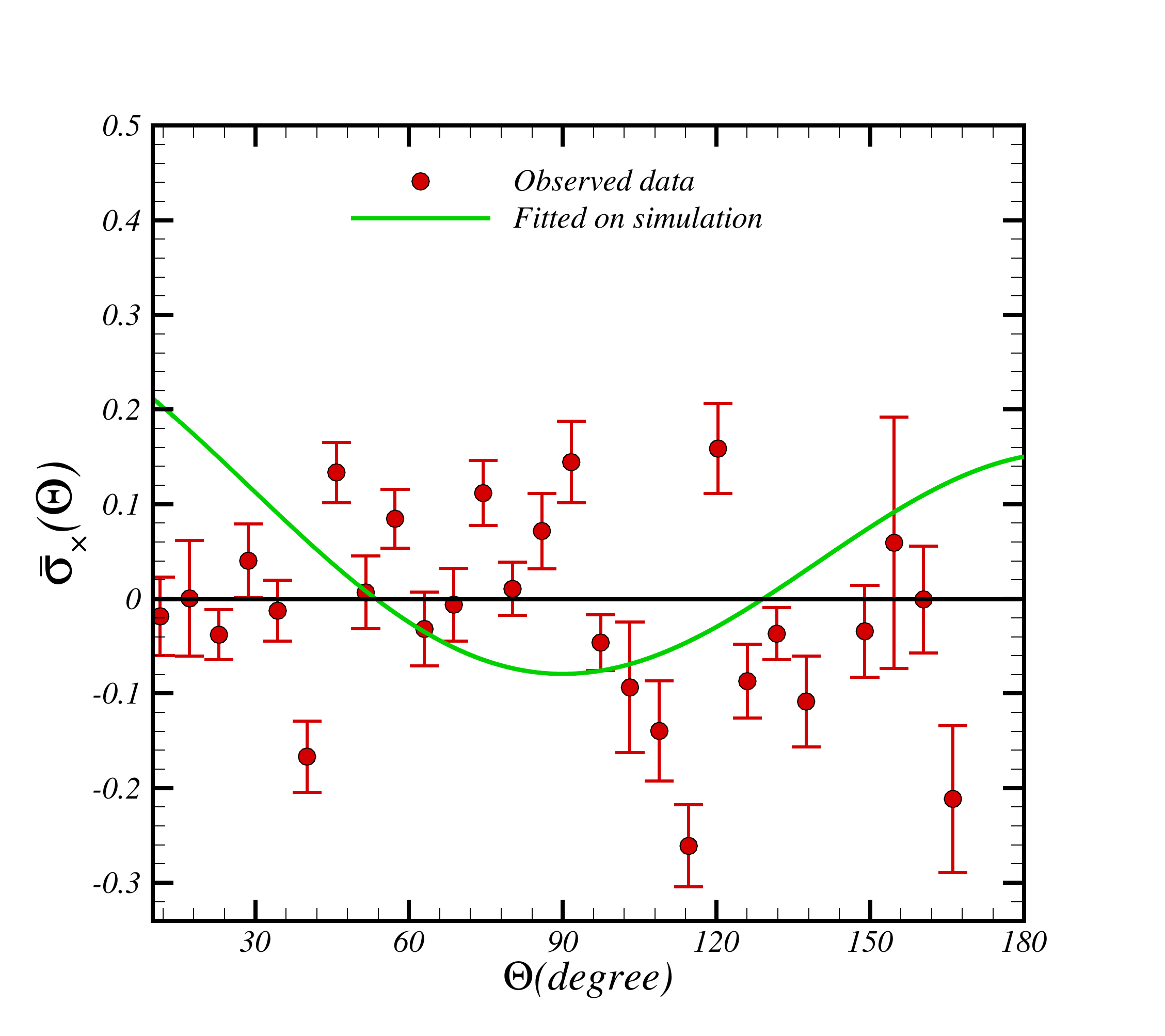}
\caption{The $\bar{\sigma}_{\times}$ as a function of $\Theta$ for 20 $PTR$s observed in the PPTA project (listed in Table \ref{tab:real}). The green line is associated with fitting on simulation.}
 \label{fig:sigmareal}
 \end{figure}

In Fig. \ref{fig:sigmareal}, we indicate $\bar{\sigma}_{\times}$ as  function of $\Theta$ for 20 MSPs observed in the PPTA project (listed in Table \ref{tab:real}). We have not obtained an obvious quadrupolar signature for the mentioned observed series due to the high value of rms, short length in the size of the data, unresolved foreground contamination, and systematic noise. In the next subsection, we will go through the finding upper limit for the amplitude of the probable GWB superimposed in observed $PTR$s.

\subsection{Upper bound on GWB amplitude}

Multifractal assessment of individual $PTR$s series is not  adequate to make a decision on the  significance of the stochastic GWB. Therefore, inspired by the unique signature of the GWB, i.e. the quadrupolar feature induced on the spatial correlation function of $PTR$s fluctuations, we apply multifractal cross-correlation analysis. This is a generalized function including spatial-temporal cross-correlation function and has some novelties compared to the standard spatial cross-correlation analysis. Our algorithm is a proper method for denoising and detrending.

The irregular MF-DXA applied to the observed irregular $PTR$s did not yield reliable results for detecting GWB due to the limited size and low S/N of the data. Different criteria introduced in this paper will enable us to detect the footprint of possible GWs with a future generation of surveys with high-S/N observations. Now we turn to assigning an upper bound on probable GWB amplitude.

Previous studies have mainly considered a model for the power spectrum of the {\it PTR} signal modulated by GWB, including the amplitude  and scaling exponent of GWB. According to priors associated with the model parameters, the Bayesian method has been adopted (\cite{lind13,Shannon15} and references therein). In our approach, we proceed with our strategies for searching the GWB (subsection \ref{startegy}).
The posterior probability function, ${\mathcal P}_{\diamond}(\mathcal{A}_{yr}|{\mathcal {D}})$, reads as:
\begin{eqnarray}\label{post1}
{\mathcal P}_{\diamond}(\mathcal{A}_{yr}|{\mathcal {D}})&\sim& {\mathcal{L}}_{\diamond}({\mathcal {D}}|\mathcal{A}_{yr}){\mathcal P}_{\diamond}(\mathcal{A}_{yr})\nonumber\\
&=& \langle\delta_D(\mathcal{A}_{yr}-\Phi_{\mathcal {D}}(\Delta h_{\diamond})) \rangle
\end{eqnarray}
Here symbol "$\diamond$" corresponds to one of four measures proposed for determining the amplitude of the stochastic GWB, and $\delta_D$ is the Dirac delta function. The $\Phi_{\mathcal {D}}(\Delta h_{\diamond})$ represents the functional form presented in Fig. \ref{fig:strategy}. The integral form of Eq. (\ref{post1}) is given by:
\begin{eqnarray}
&&{\mathcal P}_{\diamond}(\mathcal{A}_{yr}|{\mathcal {D}})=\nonumber\\
&&\int d \Delta h'_{\diamond} \mathcal{P}(\Delta h'_{\diamond}) \delta_D(\Delta h'_{\diamond}-\Delta h_{\diamond})|\mathcal{J}|_{\Delta h'_{\diamond}=\Phi_{\mathcal {D}}^{-1}(\mathcal{A}_{yr})}
\end{eqnarray}
in which $|\mathcal{J}|$ is the Jacobian computed for $\Delta h'_{\diamond}=\Phi_{\mathcal {D}}^{-1}(\mathcal{A}_{yr})$. Finally, the upper bound on $\mathcal{A}_{yr}^{up-\diamond}$ can be determined by:
\begin{equation}\label{cl}
C.L.^{\diamond}= \int_{-\infty}^{{\mathcal{A}_{yr}^{up-{\diamond}}}}d\mathcal{A}'_{yr} \mathcal{P}_{\diamond}(\mathcal{A}'_{yr}|\mathcal{D})
\end{equation}
where $C.L.^{\diamond}$ and $\mathcal{A}_{yr}^{up-\diamond}$  are the confidence interval and upper limit associated with one of our strategies, respectively.  According to the posterior function defined by Eq. (\ref{posterior}), considering $\{\mathcal {D}\}=\{\Delta h_{\diamond}^{PTR}\}$ for a given observed pulsar called by $PTR$ and $\{\Upsilon\}=\mathcal{A}_{yr}$, we compute:

\begin{equation}
\chi^2_{PTR}(\mathcal{A}_{yr})\equiv \Delta_{PTR}^{\dag}.{\mathcal C}^{-1}_{\mathcal{A}_{yr}}.\Delta_{PTR}
\end{equation}
where $\Delta_{PTR}\equiv \left[ \Delta h^{PTR}-\langle \Delta h (\mathcal{A}_{yr})\rangle  \right]$ and ${\mathcal C}_{\mathcal{A}_{yr}}$ is the $4\times4$ covariance matrix of the  four statistical features defined by $\Delta h_1$, $\Delta h_2$, $\Delta h_3$ and $\Delta h_4$ (see Eqs. (\ref{measure11}), (\ref{measure2}), (\ref{measure3}) and (\ref{measure4})). The $\langle \Delta h({\mathcal{A}}_{yr})\rangle$ is the average of $\Delta h$ over 1000 synthetic datasets for a given $\mathcal{A}_{yr}$, where $\mathcal{A}_{yr}\in[10^{-16},10^{-14}]$ and with a step size of $5\times 10^{-16}$. According to the likelihood function, ${\mathcal{L}}({\Delta h^{PTR}}|\mathcal{A}_{yr})\sim \exp(-\chi^2(\mathcal{A}_{yr})/2)$, the $95\%$ upper bound on $\mathcal{A}_{yr}^{up}$ using the observed $PTR$s is defined by:
\begin{equation}
95\%=\int_{-\infty}^{\mathcal{A}_{yr}^{up}}d\mathcal{A}_{yr} {\mathcal{L}}({\Delta h^{PTR}}|\mathcal{A}_{yr})
\end{equation}
We report the computed upper bound for some observed pulsar timing residuals at a $95\%$ confidence level in Table \ref{tab:real}. One may note that the upper bound on  $\mathcal{A}_{yr}$ has not been reported for some observed $PTR$s. This is because, for such cases, the upper value is not in the range of $\mathcal{A}_{yr}\in[10^{-16} , 10^{-14}]$ considered in this research.  Our results are consistent with other reports \citep{Shannon15}.

\begin{table*}
\centering \setlength{\tabcolsep}{4pt} \caption{Hurst exponent,
$H$, width of singularity spectrum, $\Delta \alpha$, scaling exponent
of temporal autocorrelation, $\gamma$,  rms, total time span (TTS)
of post-fit timing residuals, and the upper limit on dimensionless
amplitude of GWB of 20 MSPs observed in PPTA project. The error-bar
corresponds to $1\sigma$ confidence interval.}
\begin{minipage}{140mm}
\begin{tabular}{c*{7}{c}r}
\hline
\begin{tabular}{c} PSR\\ Number\end{tabular} & PSR Name &  $H$ & $\Delta\alpha$ & $\gamma$ &  rms $(\mu s)$  & TTS (yr) &  $\mathcal{A}_{yr}^{up}(95\%)$ \\
\hline
\\

1
&
J0437-4715
&
0.78
$\pm$
0.03
&
0.89
$\pm$
0.06
&
-1.56
$\pm$
0.06
&
0.08
&
4.76
&
$5.0\times 10^{-15}$
\\
2
&
J0613-0200
&
0.68
$\pm$
0.06
&
1.22
$\pm$
0.04
&
-1.37
$\pm$
0.11
&
1.07
&
5.99
&
$7.0\times 10^{-15}$
\\
3
&
J0711-6830
&
0.56
$\pm$
0.10
&
1.40
$\pm$
0.08
&
-1.13
$\pm$
0.19
&
0.89
&
5.99
&
$6.0\times 10^{-15}$
\\
4
&
J1022+1001
&
0.65
$\pm$
0.06
&
1.04
$\pm$
0.04
&
-1.30
$\pm$
0.13
&
1.72
&
5.88
&
$8.5\times 10^{-15}$
\\
5
&
J1024-0719
&
0.87
$\pm$
0.03
&
1.60
$\pm$
0.03
&
-1.74
$\pm$
0.07
&
1.13
&
5.99
&
-
\\
\\

6
&
J1045-4509
&
0.84
$\pm$
0.02
&
1.29
$\pm$
0.04
&
-1.68
$\pm$
0.05
&
2.77
&
5.94
&
-
\\
7
&
J1600-3053
&
0.75
$\pm$
0.05
&
1.34
$\pm$
0.04
&
-1.50
$\pm$
0.09
&
0.68
&
5.93
&
-
\\
8
&
J1603-7202
&
0.68
$\pm$
0.04
&
1.29
$\pm$
0.05
&
-1.37
$\pm$
0.07
&
2.14
&
5.99
&
$2.5\times 10^{-15}$
\\
9
&
J1643-1224
&
0.83
$\pm$
0.04
&
0.89
$\pm$
0.02
&
-1.66
$\pm$
0.08
&
1.64
&
5.87
&
-
\\
10
&
J1713+0747
&
0.74
$\pm$
0.04
&
1.20
$\pm$
0.05
&
-1.48
$\pm$
0.09
&
0.31
&
5.71
&
$2.0\times 10^{-15}$
\\
\\

11
&
J1730-2304
&
0.60
$\pm$
0.11
&
1.79
$\pm$
0.04
&
-1.21
$\pm$
0.23
&
1.47
&
5.93
&
-
\\
12
&
J1732-5049
&
0.81
$\pm$
0.03
&
1.56
$\pm$
0.03
&
-1.62
$\pm$
0.07
&
2.22
&
5.08
&
$2.0\times 10^{-15}$
\\
13
&
J1744-1134
&
0.85
$\pm$
0.04
&
1.52
$\pm$
0.03
&
-1.70
$\pm$
0.09
&
0.32
&
5.87
&
-
\\
14
&
J1824-2452A
&
0.70
$\pm$
0.03
&
1.26
$\pm$
0.05
&
-1.40
$\pm$
0.07
&
2.44
&
5.75
&
$10.0\times 10^{-15}$
\\
15
&
J1857+0943
&
0.71
$\pm$
0.05
&
1.45
$\pm$
0.02
&
-1.42
$\pm$
0.10
&
0.84
&
5.93
&
-
\\
\\

16
&
J1909-3744
&
0.76
$\pm$
0.06
&
1.32
$\pm$
0.06
&
-1.52
$\pm$
0.11
&
0.13
&
5.75
&
$6.0\times 10^{-15}$
\\
17
&
J1939+2134
&
0.80
$\pm$
0.02
&
1.25
$\pm$
0.02
&
-1.61
$\pm$
0.04
&
0.68
&
5.88
&
-
\\
18
&
J2124-3358
&
0.65
$\pm$
0.07
&
1.23
$\pm$
0.04
&
-1.30
$\pm$
0.13
&
1.90
&
5.99
&
$6.0\times 10^{-15}$
\\
19
&
J2129-5721
&
0.66
$\pm$
0.07
&
1.54
$\pm$
0.04
&
-1.32
$\pm$
0.13
&
0.80
&
5.86
&
$7.0\times 10^{-15}$
\\
20
&
J2145-0750
&
0.69
$\pm$
0.06
&
1.29
$\pm$
0.05
&
-1.38
$\pm$
0.11
&
0.78
&
5.99
&
-
\\

\\

\\
\hline
\end{tabular}
\end{minipage}
\label{tab:real}
\end{table*}

\section{Summary and conclusion}\label{sec:discussion}
The $PTR$ is a good indicator to examine relevant physical phenomena from the interior of pulsars, as well as cosmological events.
In spite of high stability in some types of pulsars, $PTR$s are classified as stochastic processes due to superimposed unknown trends and noises. The GWs produced by either primordial or late events affect the $PTR$s. Therefore, quantifying the fluctuations of $PTR$s can be a proper measure for GW detection.

In this paper, for the first time, we utilized a multifractal approach in order to examine the statistical properties of synthetic and observed $PTR$s affected by trends and noises. In the presence of trends and unknown noises, only robust methods are able to recover the correct multifractal nature of underlying series. In this research, we used MF-DFA, MF-DMA, and MF-DXA modified by the preprocessors, so-called AD or SVD algorithms. The pulsar timing observations are unevenly sampled datasets.  To mitigate this property, we modified some internal parts of the multifractal analysis and proposed the irregular MF-DXA method and examined its accuracy. Our results demonstrated that computed scaling exponents for anticorrelated and long-range-correlated irregular signals are consistent with the expectations.

We used synthetic $PTR$s simulated by the TEMPO2 pulsar timing package. A template proposed by  \cite{Hobbs_Tempo2}  was used to take into account the contribution of GWs.
We simulated 1000 synthetic $PTR$s, and the MF-DFA, MF-DMA, and MF-DXA methods were implemented on the simulated series. Our results demonstrated that the ensemble average of the Hurst exponent of the simulated data is $\langle H\rangle=0.51\pm0.02$, confirming that the pure $PTR$s belong to monofractal uncorrelated stationary processes. There is no crossover in fluctuation functions versus scale determined by MF-DFA and MF-DMA (Fig. \ref{fspure}). Adding mock GWB signal on pure $PTR$s leads to crossovers in the log-log plot of $\mathcal{F}_2$ as a function of $s$ as indicated in Fig. \ref{fspure1}. To examine the scaling behavior of  $PTR$s induced by GWs, we carried out either the SVD or AD method on the data. We found that SVD can remove the crossover on fluctuation function for any $q$.
The time scale for crossover depends on the intensity of the GW signal. In the presence of GWs, $PTR$s belong to a multifractal process due to the $q$-dependency of the generalized Hurst exponent, $h(q)$, (Fig. \ref{hqgw}). Therefore, we were able to classify the mentioned data in the universal class of the multifractal process. The value of multifractality increased by increasing the intensity of GWs.

Various components of a recorded {\it PTR} may behave as a scaling fluctuation. Therefore, applying a multifractal algorithm on individual {\it PTR}s may give spurious results in exploring  GWs. We relied on quadrupolar structure associated with the impact of GWB on the spatial cross-correlation of $PTR$s. We carried out cross-correlation analysis by the irregular MF-DXA introduced in this paper on all available $PTR$s distributed in all directions. To this end, we defined a new cross-correlation function (Eq. (\ref{eq:newsigma})) and accordingly, we computed the ensemble average of $\langle \sigma_{\times}(\Theta_{ab})\rangle_{\rm pair}$ for all synthetic $PTR$s as a function of separation angle, $\Theta$. We obtained  an analogous behavior as a quadrupolar signature in $\bar{\sigma}_{\times}$.  According to a model for GWB, obviously, the temporal part must be independent from the separation angle of the $PTR$ pairs affected by isotropic GWB, while the amplitude of cross-correlation defined by the DXA method illustrates the {\it Hellings and Downs} curve (Fig. \ref{fig:hdcca}) similar to the usual spacial crosscorrelation.

We proposed four criteria to quantify the footprint of GWs on pulsar timing residuals. Comparing the $y$-intercept of fluctuation functions with the one computed for pure $PTR$s is our first measure. The second measure is devoted to the generalized Hurst exponent with the one computed for pure $PTR$s.
Comparison between $h(q)$ and the generalized Hurst exponent computed for the Gaussian signal is the third criterion.
The fourth criterion corresponds to the width of the singularity spectrum.

The strategy for GWB detection in observations is as follows. After removing foreground and systematic noise by applying either SVD or AD on datasets, cleaned data that are associated with the dominant part of the signal (the trend part) will be used as input for irregular MF-DXA. Having observed relevant features for GWB on {\it PTR}s, irregular MF-DFA or irregular MF-DMA methods are applied exclusively. The type of superimposed GWs can be recognized by determining the Hurst exponent. Finally, the dimensionless amplitude of expected GWB ($\mathcal{A}_{yr}$) can be determined by inserting relevant quantities extracted by our four measures given by  Eqs. (\ref{measure11}), (\ref{measure2}), (\ref{measure3}) and (\ref{measure4}) for a given $\zeta$ and rms of white noise determined in observations.

There is a crossover in the log-log plot of fluctuation function versus window length of observed $PTR$s. %Instability of atomic time for mentioned observations may be a source of  cross-over.
For $s<s_{\times}$ and $s> s_{\times}$, the exponents $h(q=2)$ are $h(2)\in [1.03,1.82]$ and $h(2)\in[0.07,1.55]$, respectively.  After applying SVD, the corresponding Hurst exponent is $H\in [0.56,0.87]$.

The $q$-dependency of $h(q)$ confirmed that all observed MSPs behave as multifractal fields. The relevant exponents for observed MSPs have  been reported in Table \ref{tab:real}. The source of multifractality is mainly the correlation in small and large scales and is a universal property of all observed pulsars examined in this paper. The contribution of red-noise model indicated the extra multifractality on observed MSPs. Consequently, the degree of multifractality reported for PPTA data sets is the upper value, and a part of this value is associated with the noise model.

To infer the statistical significance of the GWB impact on the $PTR$s, we computed $\bar{\sigma}_{\times} (\Theta)$ for 20 MSPs observed in the PPTA project. Due to a high value of rms and  a short length in the size of recorded data, we have not found a quadrupolar signature. Thereafter, we computed  the upper bound for PSRs reported in Table \ref{tab:real}.

Final remarks are as follows. The observed $PTR$s are affected by noises classifying into intrinsic and extrinsic categories \cite{hobbs2006pulsar,caballero2016noise}. Reliable statistical models for noise and signal were introduced. The shuffling procedure and its evaluation by multifractal detrended analysis can also be implemented in TEMPO2 and other subroutines for simulation of $PTR$s. It could be interesting to simulate various kinds of GWs and to consider timing noise. Evaluation of different noise models and sensitivity to frequency is beyond the scope of this paper and will be considered  elsewhere.

\acknowledgements The authors thank M. Farhang for her useful comments on the manuscript. Also, the authors appreciate R. Monadi for his useful discussion.  We also appreciate W. Coles for his comments on noise models in pulsar timing residual data sets. Thanks to the anonymous referee for the very extremely useful comments and for helping us to improve this paper extensively. SMSM is grateful to the HECAP section of ICTP, where some parts of this work have been finalized.

%\clearpage
%\bibliographystyle{apj}
%\bibliography{apj-jour,biball}

%\bibliography{mybib}

\end{document}